\documentclass[twocolumn, twocolappendix]{aastex631}

\received{}
\revised{}
\accepted{}

\submitjournal{ApJ}

\shorttitle{{\jwst}-TST Proper Motions I: NIRISS Calibration and LMC Kinematics}

\shortauthors{Libralato et al.}

\usepackage{xspace}
\usepackage{amsmath}
\usepackage{rotating}
\usepackage{threeparttable}
\usepackage{enumerate}
\usepackage{color}
\usepackage{soul}

\newcommand{\hstfull}{\textit{Hubble Space Telescope}\xspace}
\newcommand{\hst}{\textit{HST}\xspace}

\newcommand{\jwst}{\textit{JWST}\xspace}
\newcommand{\gaia}{\textit{Gaia}\xspace}
\newcommand{\qfit}{\texttt{QFIT}\xspace}
\newcommand{\radxs}{\texttt{RADXS}\xspace}

\begin{document}

\title{{\jwst}-TST Proper Motions:\\ I.~High-Precision NIRISS Calibration and Large Magellanic Cloud Kinematics}

\correspondingauthor{Mattia Libralato}
\email{libra@stsci.edu}

\suppressAffiliations

\author[0000-0001-9673-7397]{Mattia Libralato}
\affil{AURA for the European Space Agency (ESA), Space Telescope Science Institute, 3700 San Martin Drive, Baltimore, MD 21218, USA}

\author[0000-0003-3858-637X]{Andrea Bellini}
\affil{Space Telescope Science Institute 3700 San Martin Drive, Baltimore, MD 21218, USA}

\author[0000-0001-7827-7825]{Roeland P.~van der Marel}
\affil{Space Telescope Science Institute 3700 San Martin Drive, Baltimore, MD 21218, USA}
\affil{Center for Astrophysical Sciences, The William H. Miller III Department of Physics \& Astronomy, Johns Hopkins University, Baltimore, MD 21218, USA}

\author[0000-0003-2861-3995]{Jay Anderson}
\affil{Space Telescope Science Institute 3700 San Martin Drive, Baltimore, MD 21218, USA}

\author[0000-0001-8368-0221]{Sangmo Tony Sohn}
\affil{Space Telescope Science Institute 3700 San Martin Drive, Baltimore, MD 21218, USA}

\author[0000-0002-1343-134X]{Laura L. Watkins}
\affil{AURA for the European Space Agency (ESA), Space Telescope Science Institute, 3700 San Martin Drive, Baltimore, MD 21218, USA}

\author[0000-0001-8703-7751]{Lili Alderson}
\affil{University of Bristol, HH Wills Physics Laboratory, Tyndall Avenue, Bristol, UK}

\author[0000-0002-0832-710X]{Natalie Allen}\altaffiliation{NSF Graduate Research Fellow}
\affil{Center for Astrophysical Sciences, The William H. Miller III Department of Physics \& Astronomy, Johns Hopkins University, Baltimore, MD 21218, USA}

\author[0000-0003-4003-8348]{Mark Clampin}
\affil{Astrophysics Division, Science Mission Directorate, NASA Headquarters, 300 E Street SW, Washington, DC 20546, USA}

\author[0000-0002-5322-2315]{Ana Glidden}
\affil{Department of Earth, Atmospheric and Planetary Sciences, Massachusetts Institute of Technology, Cambridge, MA 02139, USA}
\affil{Kavli Institute for Astrophysics and Space Research, Massachusetts Institute of Technology, Cambridge, MA 02139, USA}

\author[0000-0002-8515-7204]{Jayesh Goyal}
\affil{School of Earth and Planetary Sciences (SEPS), National Institute of Science Education and Research (NISER), HBNI, Odisha, India}

\author[0000-0002-9803-8255]{Kielan Hoch}\altaffiliation{Giacconi Fellow}
\affil{Space Telescope Science Institute 3700 San Martin Drive, Baltimore, MD 21218, USA}

\author[0000-0001-5732-8531]{Jingcheng Huang}
\affil{Department of Earth, Atmospheric and Planetary Sciences, Massachusetts Institute of Technology, Cambridge, MA 02139, USA}

\author[0000-0003-2769-0438]{Jens Kammerer}
\affil{Space Telescope Science Institute 3700 San Martin Drive, Baltimore, MD 21218, USA}

\author[0000-0002-8507-1304]{Nikole K. Lewis}
\affil{Department of Astronomy and Carl Sagan Institute, Cornell University, 122 Sciences Drive, Ithaca, NY 14853, USA}

\author[0000-0003-0525-9647]{Zifan Lin}
\affil{Department of Earth, Atmospheric and Planetary Sciences, Massachusetts Institute of Technology, Cambridge, MA 02139, USA}

\author[0000-0002-2508-9211]{Douglas Long}
\affil{Space Telescope Science Institute 3700 San Martin Drive, Baltimore, MD 21218, USA}

\author[0000-0002-2457-272X]{Dana Louie}\altaffiliation{NASA Postdoctoral Program Fellow}
\affil{NASA Goddard Space Flight Center, 8800 Greenbelt Rd, Greenbelt, MD 20771, USA}

\author[0000-0003-4816-3469]{Ryan J. MacDonald}\altaffiliation{NHFP Sagan Fellow}
\affil{Department of Astronomy, University of Michigan, 1085 S. University Ave., Ann Arbor, MI 48109, USA}

\author{Matt Mountain}
\affil{Association of Universities for Research in Astronomy, 1331 Pennsylvania Avenue NW Suite 1475, Washington, DC 20004, USA}

\author[0000-0003-2314-3453]{Maria Pe{\~n}a-Guerrero}
\affil{Space Telescope Science Institute 3700 San Martin Drive, Baltimore, MD 21218, USA}

\author[0000-0002-3191-8151]{Marshall D. Perrin}
\affil{Space Telescope Science Institute 3700 San Martin Drive, Baltimore, MD 21218, USA}

\author[0000-0003-3818-408X]{Laurent Pueyo}
\affil{Center for Astrophysical Sciences, The William H. Miller III Department of Physics \& Astronomy, Johns Hopkins University, Baltimore, MD 21218, USA}
\affil{Space Telescope Science Institute 3700 San Martin Drive, Baltimore, MD 21218, USA}

\author[0000-0002-4388-6417]{Isabel Rebollido}
\affil{Centro de Astrobiolog\'ia (CAB, CSIC-INTA), ESAC Campus Camino Bajo del Castillo, s/n, Villanueva de la Ca\~nada, E-28692 Madrid, Spain}

\author[0000-0003-4203-9715]{Emily Rickman}
\affil{AURA for the European Space Agency (ESA), Space Telescope Science Institute, 3700 San Martin Drive, Baltimore, MD 21218, USA}

\author[0000-0002-6892-6948]{Sara Seager}
\affil{Department of Earth, Atmospheric and Planetary Sciences, Massachusetts Institute of Technology, Cambridge, MA 02139, USA}
\affil{Department of Physics and Kavli Institute for Astrophysics and Space Research, Massachusetts Institute of Technology, Cambridge, MA 02139, USA}
\affil{Department of Aeronautics and Astronautics, MIT, 77 Massachusetts Avenue, Cambridge, MA 02139, USA}

\author[0000-0002-7352-7941]{Kevin B. Stevenson}
\affil{Johns Hopkins APL, 11100 Johns Hopkins Rd, Laurel, MD 20723, USA}

\author[0000-0003-3305-6281]{Jeff A. Valenti}
\affil{Space Telescope Science Institute 3700 San Martin Drive, Baltimore, MD 21218, USA}

\author[0000-0002-2643-6836]{Daniel Valentine}
\affil{University of Bristol, HH Wills Physics Laboratory, Tyndall Avenue, Bristol, UK}

\author[0000-0003-4328-3867]{Hannah R. Wakeford}
\affil{University of Bristol, HH Wills Physics Laboratory, Tyndall Avenue, Bristol, UK}

\collaboration{30}{}

\begin{abstract}
We develop and disseminate effective point-spread functions and geometric-distortion solutions for high-precision astrometry and photometry with the \jwst NIRISS instrument. We correct field dependencies and detector effects, and assess the quality and the temporal stability of the calibrations. As a scientific application and validation, we study the proper motion (PM) kinematics of stars in the \jwst calibration field near the Large Magellanic Cloud (LMC) center, comparing to a first-epoch \hstfull (\hst) archival catalog with a 16-yr baseline. For stars with ${\it G} \sim 20$, the median PM uncertainty is $\sim$13 $\mu$as yr$^{-1}$ (3.1 km s$^{-1}$), better than \gaia DR3 typically achieves for its very best-measured stars. We kinematically detect the known star cluster OGLE-CL LMC 407, measure its absolute PM for the first time, and show how this differs from other LMC populations. The inferred cluster dispersion sets an upper limit of $24$ $\mu$as yr$^{-1}$ ($5.6$ km s$^{-1}$) on systematic uncertainties. Red-giant-branch stars have a velocity dispersion of $33.8 \pm 0.6$ km s$^{-1}$, while younger blue populations have a narrower velocity distribution, but with a significant kinematical substructure. We discuss how this relates to the larger velocity dispersions inferred from \gaia DR3. These results establish \jwst as capable of state-of-the-art astrometry, building on the extensive legacy of \hst. This is the first paper in a series by our \jwst Telescope Scientist Team (TST), in which we will use Guaranteed Time Observations to study the PM kinematics of various stellar systems in the Local Group.
\end{abstract}

\keywords{astrometry -- photometry -- proper motions -- stellar clusters -- Large Magellanic Cloud}

\section{Introduction}\label{sec:intro}

\jwst's primary objectives \citep{2023arXiv230404869G} are to study the first galaxies in the Universe, shed light on the birth of stars and planets, and study/explore other worlds, but its cutting-edge design and state-of-the-art detectors are poised to revolutionize astronomy over a much broader context.

From simple source centroiding to large-area image mosaics, a multifaceted variety of scientific applications with \jwst, from stars and exoplanets in the Solar neighborhood to structures in the Local Group, requires accurate and/or precise astrometry and photometry that may (and often do) rely on better calibrations than those generally offered by the official calibration pipeline (e.g., distortion uncertainty not to exceed 5 mas per coordinate in all detectors; see Anderson 2016\footnote{Document \href{https://www.stsci.edu/files/live/sites/www/files/home/jwst/documentation/technical-documents/_documents/JWST-STScI-005361.pdf}{JWST-STScI-005361}, ``Verification of Plan to Solve for the Distortion Solution''.}).

Here we begin to extend the techniques that made the \hstfull (\hst) and its cameras among the best resources for high-precision photometry and astrometry \citep[e.g.,][]{2000PASP..112.1360A,2004acs..rept...15A,2006acs..rept....1A} to \jwst's Near Infrared Imager and Slitless Spectrograph (NIRISS) camera \citep{2012SPIE.8442E..2RD}. NIRISS provides various observing modes, including imaging, and---because of its detector size and filter properties---it can be considered a complementary channel to the Near InfaRed Camera (NIRCam). In Cycle 1, NIRISS imaging was offered only for parallel observations but, starting from Cycle 2, it can also be used as the primary instrument. So far, most of the attention from the imaging community has been on NIRCam \citep{2022MNRAS.517..484N,2022arXiv221203256G}, while many of the astrometric and photometric capabilities of NIRISS remain to be demonstrated.

We present careful procedures with which we modeled the effective point-spread function \citep[ePSF\footnote{We refer to the PSF discussed and solved for here as the ``effective'' PSF, in that it represents the ``instrumental'' PSF that impinges on the detector convolved with an image pixel.},][]{2000PASP..112.1360A} and geometric-distortion (GD) correction of the NIRISS detector for all 12 of its imaging filters. We also briefly discuss the quality and stability of our models over the limited temporal baseline available ($\sim$2 months). Finally, we demonstrate the scientific potential of our methods and calibrations  by computing proper motions (PMs) of stars in a field in the Large Magellanic Cloud (LMC).

This paper is the first of many to be written by the \textit{JWST} Telescope Scientist Team\footnote{\href{https://www.stsci.edu/~marel/jwsttelsciteam.html}{https://www.stsci.edu/$\sim$marel/jwsttelsciteam.html}.}. The team, led by M.~Mountain, was convened in 2002, following a competitive NASA selection process. In addition to providing scientific support for observatory development through launch and commissioning \citep[e.g.,][]{2012OptEn..51a1005G,2014SPIE.9143E..09P,2015JATIS...1a4004G,2016SPIE.9904E..0FP,2018SPIE10698E..3NL}, the team was awarded 210 hours of Guaranteed Time Observer (GTO) time. This time is being used over the first three {\jwst} observing cycles for studies in three different subject areas: (a) Transiting Exoplanet Spectroscopy (lead: N. Lewis); (b) High Contrast Imaging of Exoplanetary Systems (lead: M. Perrin); and (c) Local Group Proper Motion Science (lead: R. van der Marel). A common theme of these investigations is the desire to pursue and demonstrate science for the astronomical community at the limits of what is made possible by the exquisite optics and stability of {\jwst}. The present paper is part of our work on Local Group Proper Motion Science, which focuses on the Galactic Center and selected satellite companions of both the Milky Way and Andromeda galaxies. We present here methods, calibrations, and scientific validations using archival data outside our GTO program that are essential for our subsequent analysis of new GTO data obtained with the NIRISS instrument.

The paper is organized as follows. Section~\ref{sec:data} describes the data sets used in our analyses. Sections~\ref{sec:psf} and \ref{sec:gd} describe the methodology used to obtain accurate ePSF models and GD corrections, respectively, and include analyses of their quality and temporal stability. Section~\ref{sec:science} presents a scientific application focused on measuring the PM kinematics of various populations in the \jwst calibration field in the Large Magellanic Cloud (LMC). Finally, Section~\ref{sec:release} presents the ePSF models and GD corrections that we release to the community with this paper. Section~\ref{sec:conc} summarizes our main conclusions.

\section{Data sets}\label{sec:data}

We mostly made use of data from Commissioning program 1086 (PI: Martel) that observed the \jwst calibration field in the LMC centered at $(\alpha,\delta) = (80.5,-69.5)$ deg \citep{2021jwst.rept.7716A} on 2022 May 1. The data set consists of 9 pointings over a 3$\times$3 dither pattern and two images per pointing for each of the 12 NIRISS filters. Each exposure was taken with the {\mbox{NISRAPID}}\footnote{\href{https://jwst-docs.stsci.edu/jwst-near-infrared-imager-and-slitless-spectrograph/niriss-instrumentation/niriss-detector-overview/niriss-detector-readout-patterns}{https://jwst-docs.stsci.edu/jwst-near-infrared-imager-and-slitless-spectrograph/niriss-instrumentation/niriss-detector-overview/niriss-detector-readout-patterns}.} readout pattern, one integration, and between five and 15 groups, depending on the filter.

Our external validation and stability analyses made use of two additional data sets: the Early Release Science (ERS) program 1334 (PI: Weisz) and the Cycle-1 calibration program 1501 (PI: Sohn). The ERS program observed the globular cluster M92 through the F090W and F150W filters using the NIS readout pattern, one integration, and seven groups. These ERS NIRISS images were taken between 2022 June 20-21, as parallel observations to NIRCam primary exposures, and as such the dither pattern was driven by NIRCam's needs. The Cycle-1 images were taken on 2022 September 18 and were focused on the LMC astrometric field for calibration purposes, and consist of four exposures per filter with the same readout pattern of Commissioning program 1086. The present study requires the use of unresampled images, which we downloaded from the MAST\footnote{\href{https://mast.stsci.edu/portal/Mashup/Clients/Mast/Portal.html}{https://mast.stsci.edu/portal/Mashup/Clients/Mast/Portal.html}.} archive as level-2, \texttt{\_cal} products. The three data sets\footnote{All of the data presented in this paper were obtained from the Mikulski Archive for Space Telescopes (MAST) at the Space Telescope Science Institute. The specific observations analyzed can be accessed via \dataset[DOI: 10.17909/hv07-tg13]{https://doi.org/10.17909/hv07-tg13}.} were downloaded from the archive at different times, and they have been calibrated using different reference calibration files given the fast update rate of calibration reference files in the Calibration Reference Data System (CRDS)\footnote{\href{https://jwst-crds.stsci.edu/}{https://jwst-crds.stsci.edu/}.} during Commissioning and in the early days of Cycle-1 (see Table~\ref{tab:context} in Appendix~\ref{context}).

\section{ePSF modeling}\label{sec:psf}

The first step to achieve precise measurements of positions and fluxes for point-like sources is to obtain accurate ePSF models. The NIRISS ePSFs are undersampled below 3.8 $\mu$m, i.e., for 8 out of its 12 filters, and extra care must be taken in modeling the ePSFs to avoid systematic effects.

The construction of accurate PSFs involves an inherently circular problem: to extract a PSF from an image, we need accurate positions of stars, but we cannot measure accurate positions of stars without an accurate model of the PSF. The use of non-optimal ePSFs tends to result in positions that are affected by significant pixel-phase biases, i.e., positions that are systematically measured at a specific location with respect to the pixel boundaries regardless of where sources truly are. \citet{2000PASP..112.1360A} showed that well-dithered images can help to break this degeneracy, since dithering allows a given star to land at different locations with respect to the pixel boundaries. The use of a common reference frame, in which positions of sources measured in different dithered exposures are averaged together, also helps in mitigating pixel-phase systematics, as we can transform these averaged positions back into the raw coordinate system of each individual \texttt{\_cal} image and use them as improved positions from which to construct the ePSF models. The process is iterated to break the circularity problem mentioned above: once improved ePSF models are made, we can measure improved stellar positions in each exposure, resulting in a more precise averaged position on the common frame that can be inverted to construct even better ePSF models. The cycle is repeated until convergence is reached. We closely follow the prescriptions provided in \citet{2000PASP..112.1360A}, \citet{2016wfc..rept...12A}, and \citet{2016MNRAS.463.1780L,2021MNRAS.500.3213L}. We refer the interested readers to these papers for a more detailed description of the method, and we provide a brief overview below.

The value $P_{i,j}$ of a pixel $(i,j)$  in the vicinity of a star located at $(x_{\ast},y_{\ast})$ can be defined as:
\begin{displaymath}
  P_{i,j} = z_{\ast} \cdot \psi_{\rm E}(i-x_{\ast},j-y_{\ast}) + s_{\ast}
  \phantom{1} ,
\end{displaymath}
where  $z_{\ast}$ is the stellar flux, $s_{\ast}$ is the local sky background, and the fraction of star's light that should fall on that pixel according to the ePSF model is given by $\psi_{\rm E}(i-x_{\ast},j-y_{\ast})$. The ePSF fit allows us to obtain the stellar parameters $(x_{\ast}, y_{\ast}, z_{\ast})$. Since we are  interested in modeling the ePSF, we invert the equation above to obtain:
\begin{displaymath}
    \psi_{\rm E}(i-x_{\ast},j-y_{\ast}) = \frac{P_{i,j}-s_{\ast}}{z_{\ast}} \phantom{1} .
\end{displaymath}
The knowledge of $(x_{\ast}, y_{\ast}, z_{\ast})$ for a star makes each pixel in the star's vicinity a potential sampling of the ePSF. Using many samplings from many stars in many images that uniformly map the pixel space, we can build reliable ePSF models.

The iterative process described above starts by measuring positions and fluxes of bright, isolated stars in each individual NIRISS image. Positions were initially obtained by simply estimating the light-center of the flux distribution in the innermost pixels of each source, and fluxes were initially measured via aperture photometry. As soon as ePSF models are available, these stellar parameters are re-measured via ePSF fitting.

The same stars were then cross-identified in multiple catalogs, and their positions and fluxes were averaged together after being transformed on to a common reference-frame system (hereafter, the master frame\footnote{Orientation and scale for the master frame was set up using the \hst catalog of \citet{2021jwst.rept.7716A}, while the photometry was registered to that of a NIRISS image at the center of the dither patter; see also Sect.~\ref{sec:science}.}) by means of six-parameter linear transformations.  These averaged positions and fluxes, once transformed back into the frames of each exposure, represented the (current) best estimates of $(x_{\ast}, y_{\ast}, z_{\ast})$, which were used to collect the samplings for the ePSF reconstruction. It is at this point that a GD correction is often necessary to transform positions back and forth between different frames\footnote{If the dither is small and the distortion is small, then sometimes the differential distortion among the exposures can be ignored, but the dithers here were large.}. Without GD correction, the ePSF samplings would be systematically misplaced, thus preventing us from mitigating pixel-phase biases. For this reason, we made use of a GD correction derived following the procedure highlighted in Sect.~\ref{sec:gd}. This correction was itself improved during the iteration process.

Thanks to the availability of many thousands of stars in several dithered images, we were able to model the ePSFs using an oversampling factor of 4. We initially constrained one single ePSF model for the entire detector, but later on during the iteration process we let the ePSF model vary spatially. We found that a 5$\times$5 array of ePSF models is able to fully capture the spatial variability of the ePSF across the field of view (FoV) of NIRISS. At the beginning of each iteration, newly-derived ePSF models were used to measure improved positions and fluxes, which in turn defined an improved master frame. Convergence of the ePSF models was reached in 20 iterations, with subsequent iterations providing no significant improvements. During the last few iterations, we allowed the ePSF fitting procedure to also fit source positions instead of imposing positions from the master frame.

\begin{figure}[t!]
    \centering
    \includegraphics[width=\columnwidth]{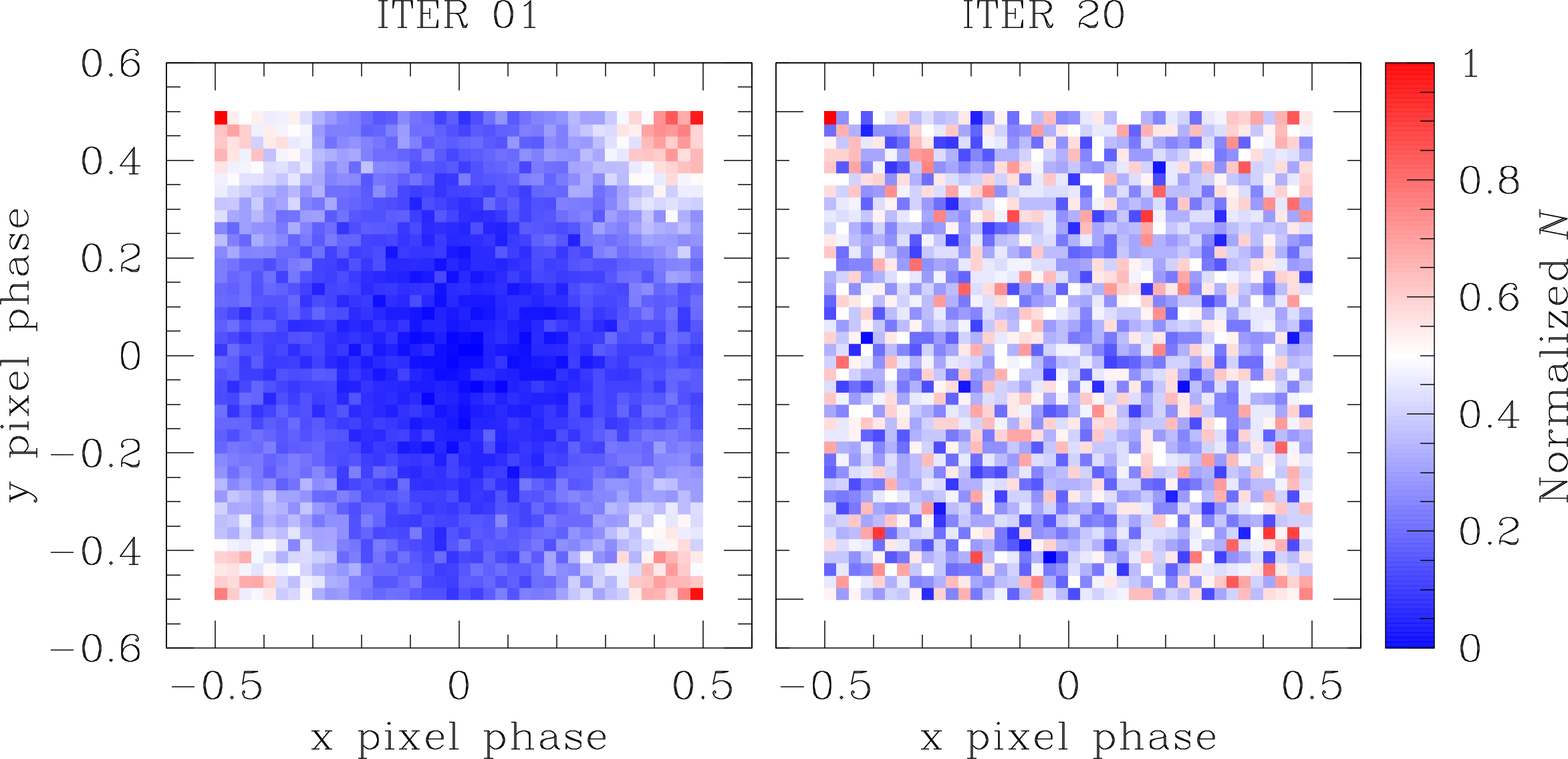}
    \caption{Density plots showing where the center of stars are measured to be with respect to the pixel boundaries in F090W images during our iteration process. The maps are color coded according to the normalized density shown in the colorbar on the right. The left panel corresponds to the first iteration, in which positions measured using photocenters are clearly biased towards the corners of pixels. The density plot on the right, obtained in our last iteration, clearly shows that the pixel-phase bias is removed when positions are estimated using precise ePSF models.}
    \label{fig:f090w_pp}
\end{figure}

\begin{figure}[t!]
    \centering
    \includegraphics[width=\columnwidth]{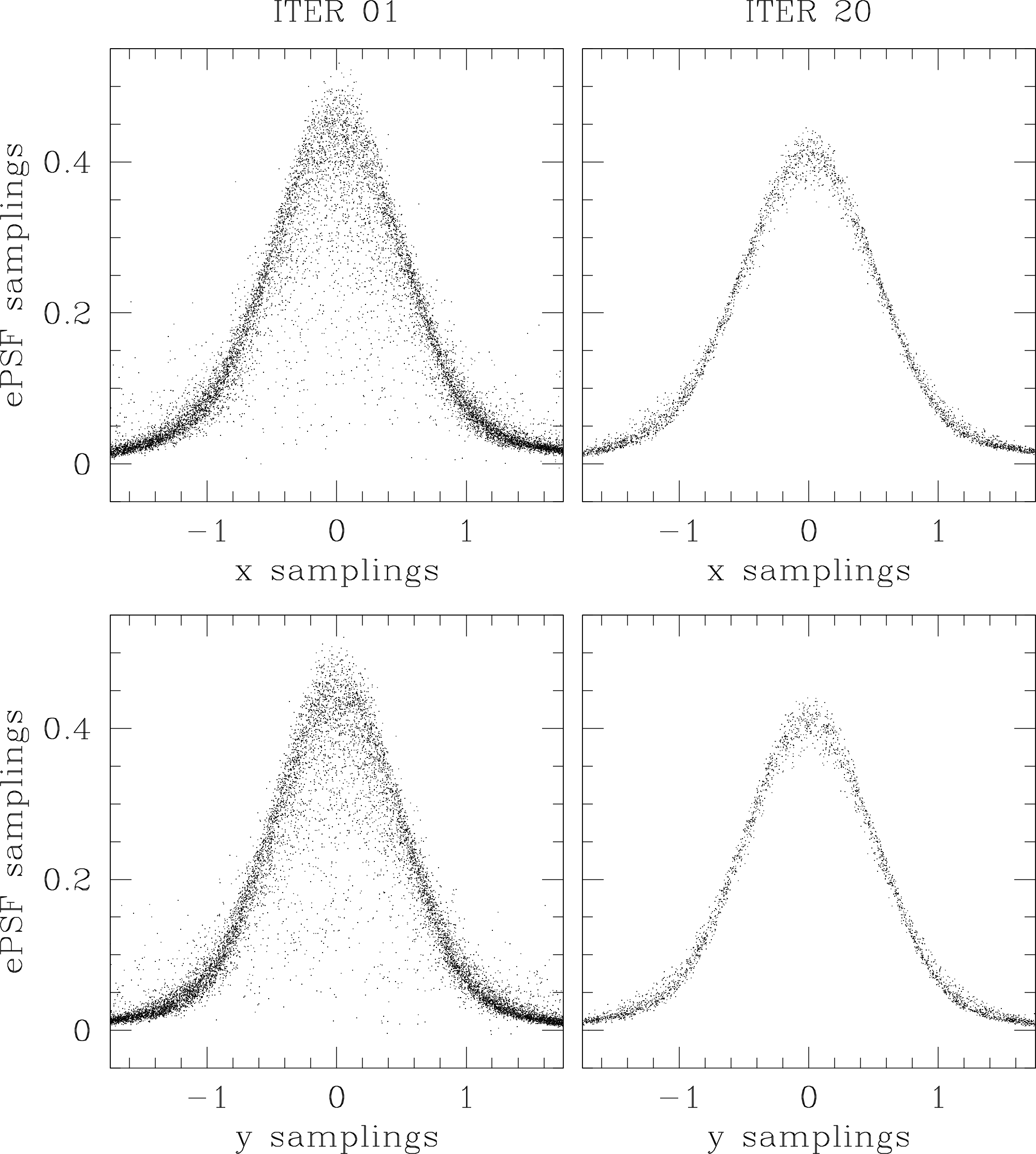}
    \caption{ePSF samplings ($\psi_{\rm E}$) with respect to the center of the ePSF placed at 0 along the $x$ (top panels) and $y$ (bottom panels) axes in a 0.02-pixel-wide strip around the center. Each point correspond to a sampling collected from a pixel of a star in an image. As in Fig.~\ref{fig:f090w_pp}, the left panels are for the first iteration, while the right panels show the final samplings at the end of our iterative process. Initially, the core of the ePSF is too sharp and not well constrained because stellar positions are not well defined. After 20 iterations, the ePSF samplings are smooth, with less scatter than before, and the shape of the ePSF core is well constrained.}
    \label{fig:f090w_sam}
%
    \centering
    \includegraphics[width=0.98\columnwidth]{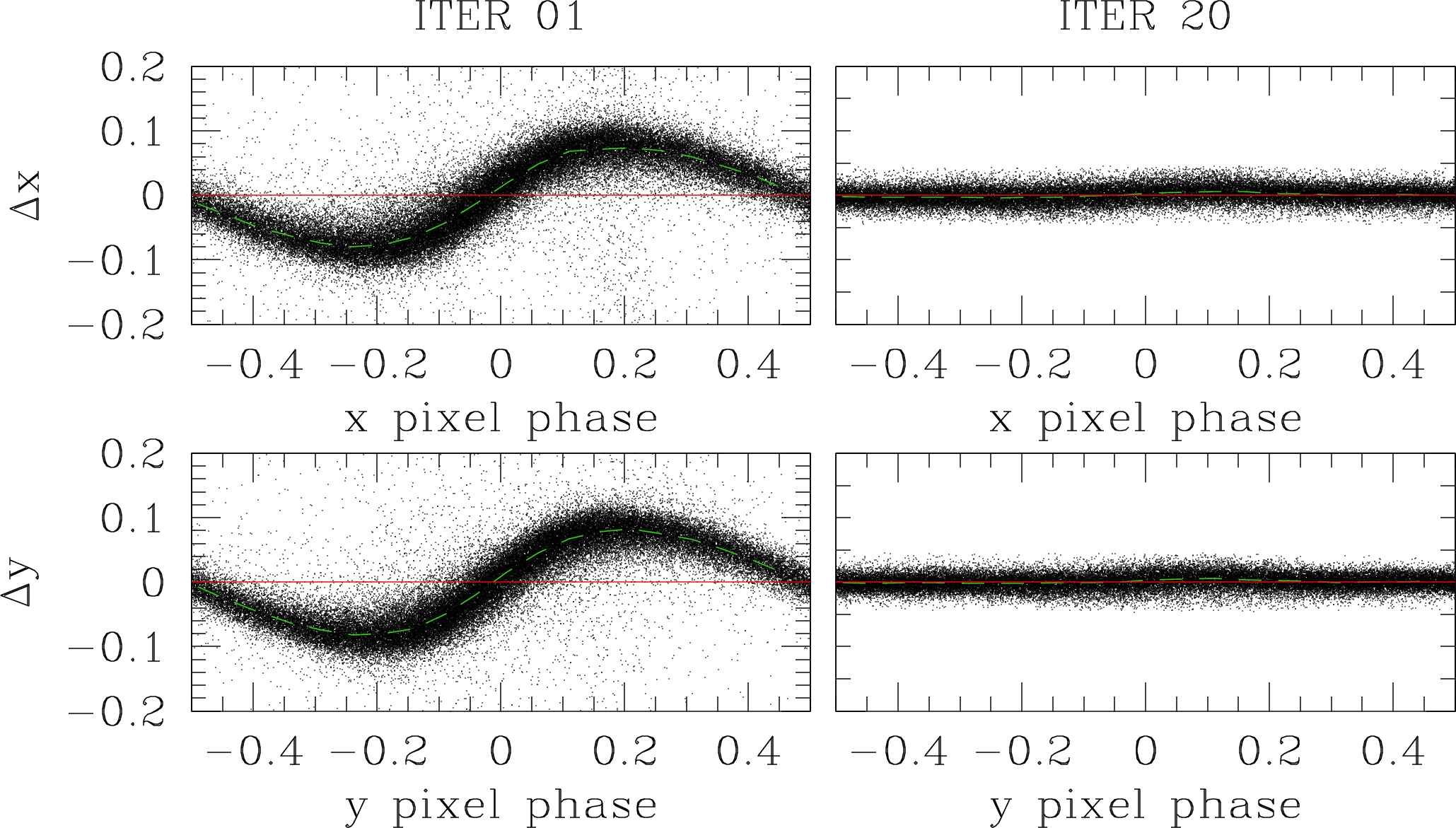}
    \caption{Initial (left) and final (right) pixel-phase errors for the $\Delta x$ (top) and $\Delta y$ (bottom) positional residuals, defined as the difference between the single-exposure, ePSF-fit positions and the master-frame averaged positions (once transformed back on to the frame of the single exposures). The red horizontal lines at 0 are used as a reference, while the green, dashed curves represent the median trend. Residuals in the final iterations are, on average, smaller than 0.01 pixels, and only show a marginal trend with pixel phase.}
    \label{fig:f090w_pperr}
\end{figure}

\begin{figure*}[t!]
    \centering
    \includegraphics[width=\textwidth]{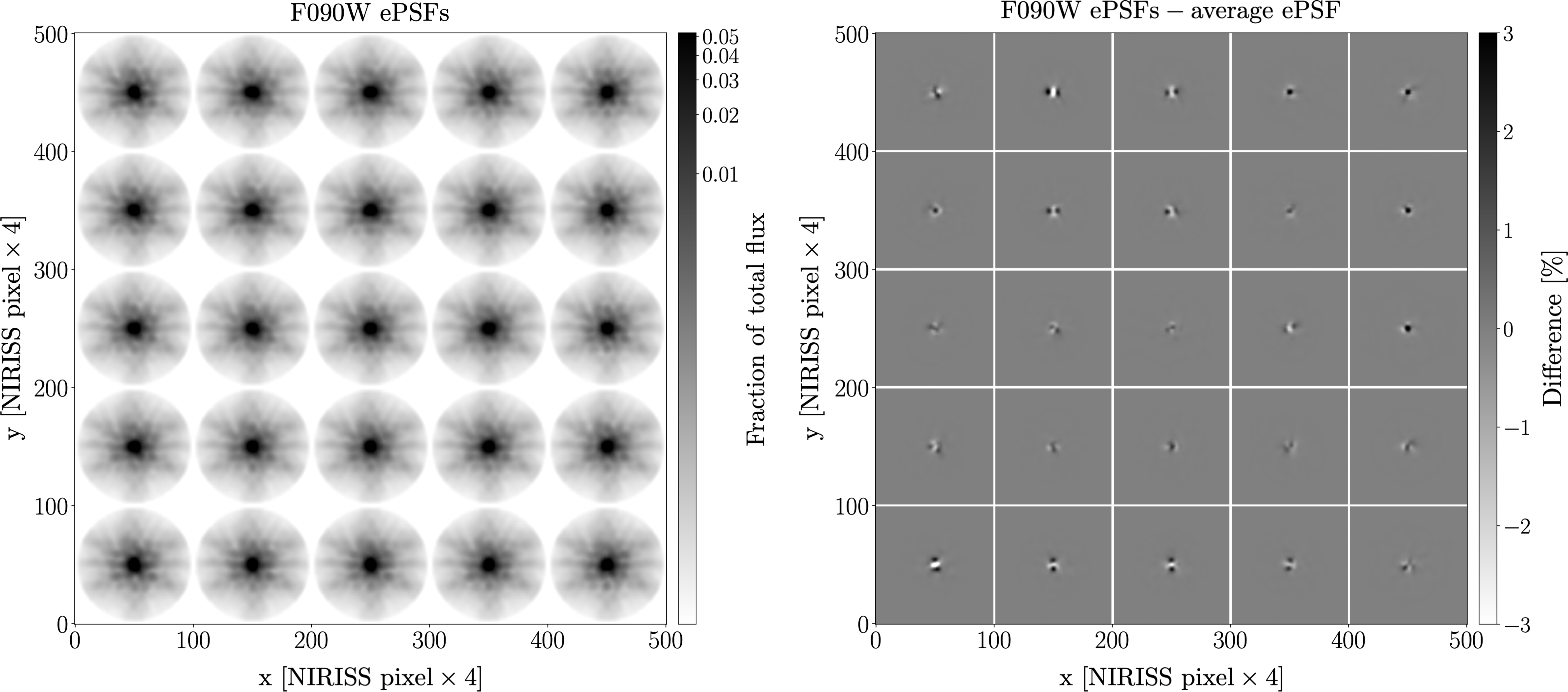}
    \caption{The left plot shows the 5$\times$5 array of ePSF models for the F090W filter in logarithmic scale. The difference between each F090W ePSF and the average ePSF is shown in the right panel in linear scale. }
    \label{fig:f090w_psf}
\end{figure*}

As discussed in \citet{2000PASP..112.1360A} and \citet{2016wfc..rept...12A}, some constraints should be applied to the ePSF models to make sure they are smooth and continuous. We achieved this by convolving the ePSFs with a series of low-pass filters that allowed less high-frequency structure at larger radii. These smoothing kernels were empirically chosen by trial and error \citep[see also][]{2016wfc..rept...12A} and are a compromise between the need to have smooth and continuous ePSF models while preserving true sharp ePSF features. The ePSF models were also normalized to sum up to 1 within a radius of 5 NIRISS pixels (chosen arbitrarily).

Figures~\ref{fig:f090w_pp}, \ref{fig:f090w_sam} and \ref{fig:f090w_pperr} illustrate the main pixel-phase issues that affect undersampled detectors and that can be minimized using a careful modeling of the ePSFs. The Figures refer to the F090W case, which is the most undersampled among the NIRISS filters. At the beginning of our iterative process: (i) stars are preferentially measured to be at the corners of the pixels, even though their true position is random within the pixels (left panel of Fig.~\ref{fig:f090w_pp}); (ii) the core of the ePSF is too sharp and not well constrained (left panels of Fig.~\ref{fig:f090w_sam}); and (iii) the positional residuals (defined as the difference between the raw positions in each exposure and those predicted by the inverted averaged master frame) show a clear trend as a function of pixel phase (left panels of Fig.~\ref{fig:f090w_pperr}). All these features are caused by the undersampled nature of the ePSF: a sharp PSF does not transfer flux linearly from pixel to pixel, and therefore a linear center of light model mis-measures positions.  After 20 iterations in which the true shape of the PSF is teased out: (i) stellar positions measured by fitting our precise ePSF models do not show any preferential location with respect to the pixel boundaries (right panel of Fig.~\ref{fig:f090w_pp}); (ii) the ePSF samplings are smooth, with less scatter than before, and the shape of the ePSF core is well constrained (right panels of Fig.~\ref{fig:f090w_sam}); and (iii) pixel-phase errors have been significantly reduced (right panels of Fig.~\ref{fig:f090w_pperr}).

To summarize, our final ePSF models are spatially-variable 5$\times$5 arrays of ePSFs covering the full NIRISS detector. Each ePSF has a size of 101$\times$101 oversampled pixels, with the ePSF centered at location (51,51), extending out 12.5 real NIRISS pixels in each direction, and it is normalized to have unity flux within a radius of 5 NIRISS pixels. The core of the ePSF models becomes progressively less sharp towards redder wavelengths (i.e., the less undersampled the ePSFs are), with the fraction of the flux in the centermost pixel of the ePSF model decreasing from $\sim$46\% (F090W filter) to $\sim$11\% (F480M filter).

The left panel of Figure~\ref{fig:f090w_psf} shows the full 5$\times$5 ePSF array for the F090W filter. The differences between each spatially-varying ePSF and the average ePSF are shown in the right panel. The spatial variation of the F090W models causes about $6\%$ of the central-pixel flux to be redistributed into the wings. The ePSFs of redder filters are found to be less affected by spatial variations, with as low as only $0.3\%$ of the central-pixel flux being redistributed in the reddest F480M filter.

In the following, we refer to these final ePSF models as the \textit{library} ePSFs. We will discuss strategies to further improve these models on an image-by-image basis in Sect.~\ref{sec:psftime}.

\subsection{Quality of our ePSFs}\label{sec:psffit}

\begin{figure*}
    \centering
    \includegraphics[width=\textwidth]{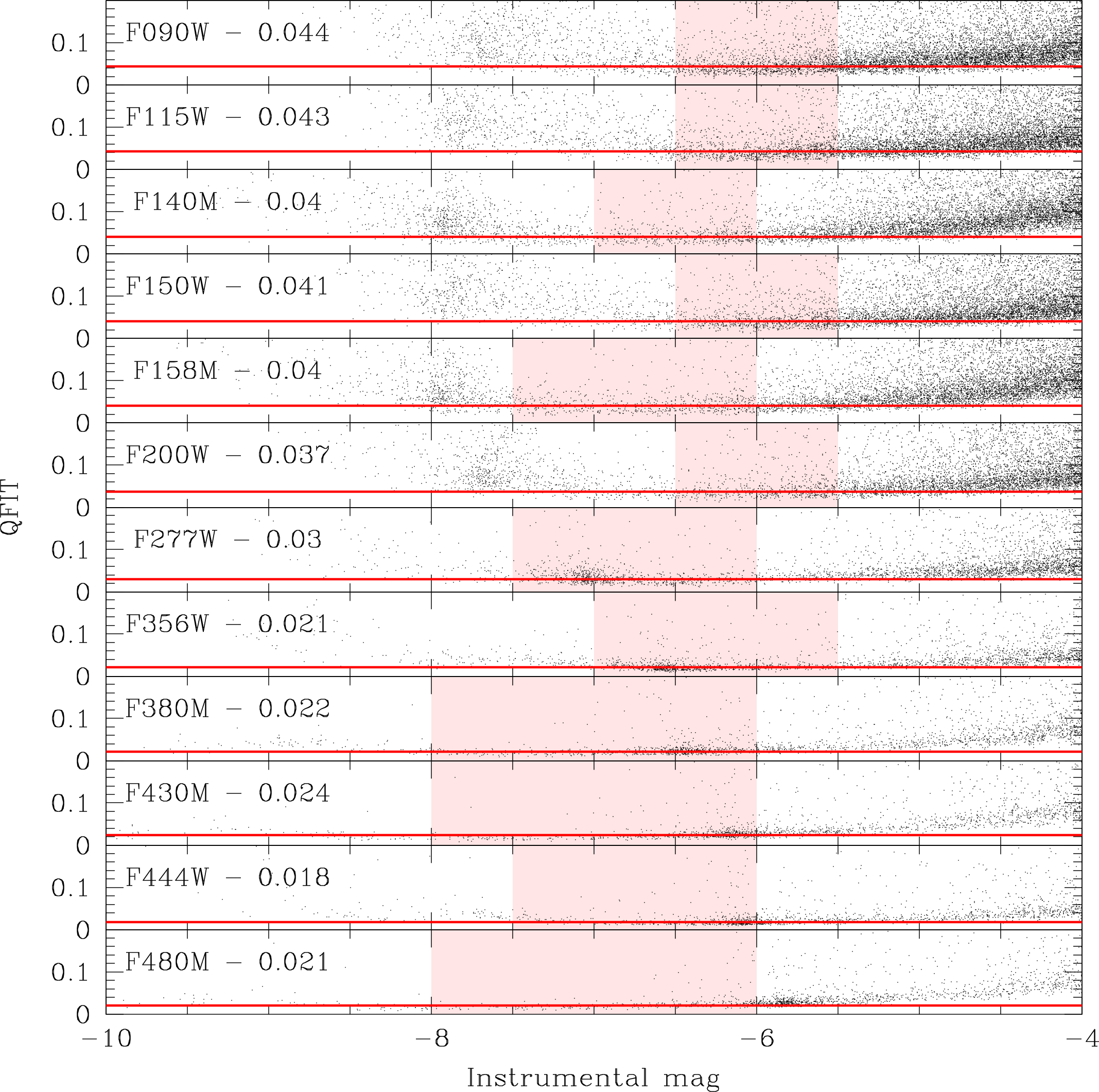}
    \caption{Examples of \qfit as a function of instrumental magnitude for one exposure in each filter obtained using the library ePSFs. The red, horizontal line is set at the median value of the \qfit (also displayed in each panel) for the stars within the pink, shaded region.}
    \label{fig:qfit_lib}
\end{figure*}

\begin{figure*}
    \centering
    \includegraphics[width=\textwidth]{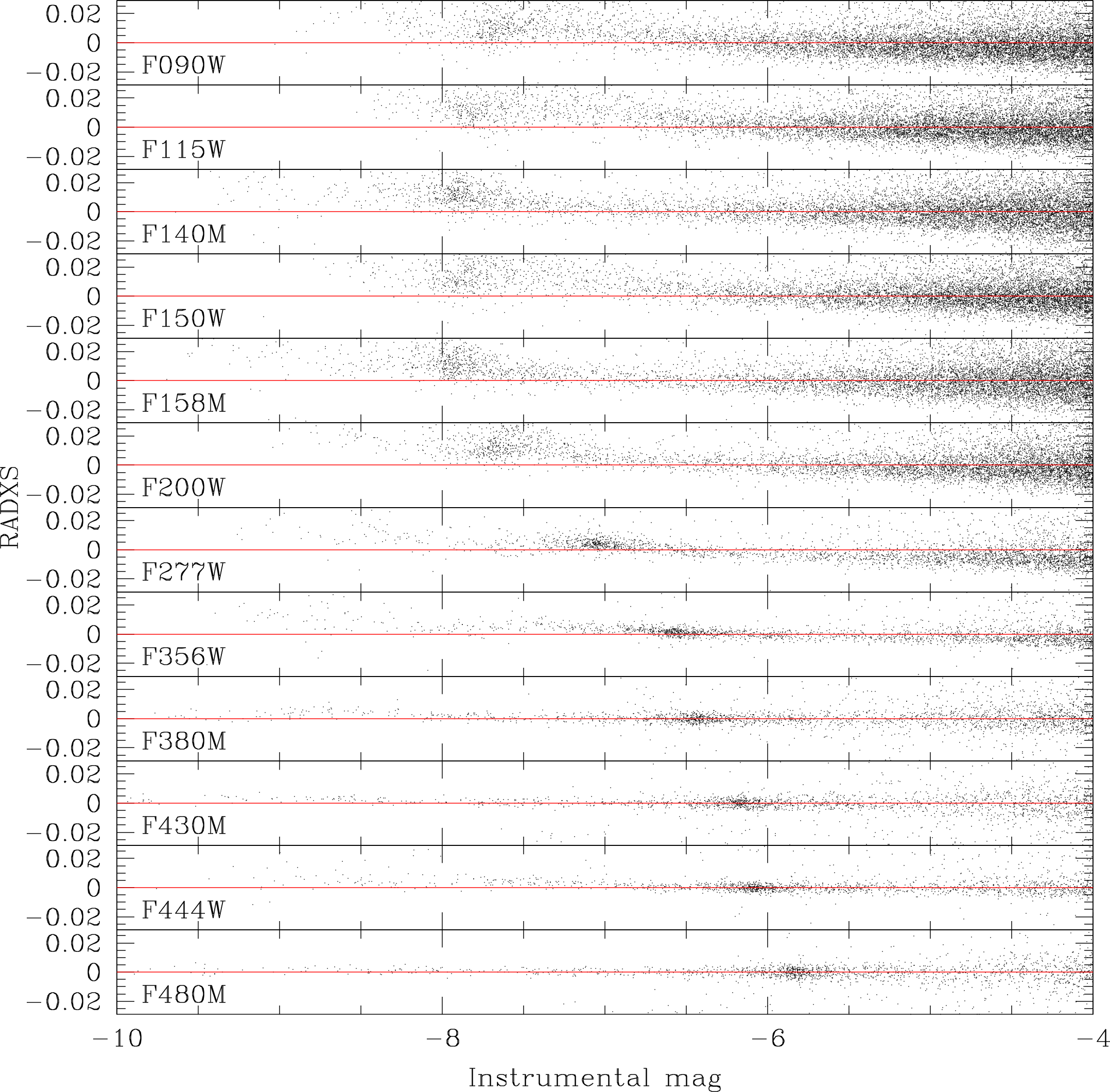}
    \caption{\radxs as a function of instrumental magnitude for one indicative exposure per filter from  the Commissioning data. As for Fig.~\ref{fig:qfit_lib}, the library ePSFs were fit to estimate the stellar parameters. The red, horizontal line is set at 0 as a reference.}
    \label{fig:radxs}
\end{figure*}

We used two diagnostics to test our library ePSF models. The first one is the ``Quality of PSF fit" or \qfit \citep{2014A&A...563A..80L}, which is defined as the absolute fractional error in the fit of a source. Well-measured stars have a \qfit close to 0, whereas increasingly poorly-measured objects have increasingly larger \qfit values. The result of this first test for one indicative image per filter from the Commissioning data set is shown in Fig.~\ref{fig:qfit_lib}. The red lines are set at the median \qfit values of the bright stars in a region where the \qfit trend is flat (shaded pink regions). These medians, reported on the left side of each panel, are always below 0.05, a value generally used as a proxy for adequate ePSF models. The \qfit increases towards fainter magnitudes due to the lower signal-to-noise ratio. For the bluest, more undersampled filters, the brightest stars also show an increase of the \qfit trend. This is likely due to a combination of saturation, non-linearity and/or brighter-fatter effects. We also do not exclude that quantum yield\footnote{Quantum yield describes the possibility, for the most energetic photons, to generate more than one electron-hole pair per detection.} could play a role for the bluest filters. We investigated the possible presence of chromatic effects in our ePSFs by comparing the \qfit values of the bluest and reddest stars with high signal to noise, and found no systematic differences. This is somewhat expected since we did not restrict our ePSF modeling to stars of specific colors.

The second diagnostic parameter is \radxs: the excess/deficiency of flux outside the core of the sources with respect to the ePSF prediction \citep{2008ApJ...678.1279B}. \radxs is a powerful tool to discriminate between point-like sources (\radxs $\sim 0$), cosmic rays or hot pixels (negative \radxs), and galaxies (positive \radxs). If our ePSF models are a good representation of the flux distribution of point-like sources outside their cores, the \radxs of stars should be close to zero. Figure~\ref{fig:radxs} presents \radxs trends as a function of instrumental magnitude\footnote{The instrumental magnitude is defined as $-2.5\log$(total surface brightness), where ``total surface brightness" is the total surface brightness (level-2 \texttt{\_cal} FITS files are in units of MJy sr$^{-1}$) under the fitted ePSF.} for the same exposures used in Fig.~\ref{fig:qfit_lib}. While the general \radxs trends are all around zero, small gradients are visible in all but the reddest filters, with \radxs preferentially positive for bright sources and preferentially negative for faint sources. We suspect this behavior is due to brighter-fatter effects \citep[e.g.,][]{2014JInst...9C3048A,2018PASP..130f5004P}, which are known to be present in the NIRISS detector\footnote{\href{https://jwst-docs.stsci.edu/jwst-near-infrared-imager-and-slitless-spectrograph/niriss-instrumentation/niriss-detector-overview/niriss-detector-performance\#NIRISSDetectorPerformance-Detectorfullwellcapacity}{https://jwst-docs.stsci.edu/jwst-near-infrared-imager-and-slitless-spectrograph/niriss-instrumentation/niriss-detector-overview/niriss-detector-performance\#NIRISSDetectorPerformance-Detectorfullwellcapacity}.}. The brighter-fatter effects causes photons that should land on pixels that are closer to saturation to end up generating a photo-electron in a neighboring pixel, resulting in bright point-like sources appearing ``fatter", i.e., broader, than fainter ones. Since our library ePSF models were constructed using stars  within specific luminosity ranges, fitting sources outside these ranges might provide poorer results due to brighter-fatter effects. The systematic trends in Fig.~\ref{fig:radxs} are more pronounced in bluer (more undersampled) filters. We find that the impact of the brighter-fatter effect on astrometry with the NIRISS data is very small, and is only relevant for the bluest, most undersampled filters. For example, Fig.~\ref{fig:f090w_pperr} shows that after 20 iterations in our ePSF-modeling procedure, the pixel-phase errors are within $\pm$0.01 pixel along the $x$ direction. The amplitude of these residuals seems not to depend on the magnitude of the star, but the patterns for bright and faint stars have a different phase. This is another piece of evidence that the ePSF of bright and faint stars is slightly different because the flux distributions for the two groups of stars differ, as one would expect from the brighter-fatter effect. A detailed analysis of this phenomenon is beyond the scope of this paper, but we want to make users aware of its presence. Flux-dependent ePSF models, similarly to what is done in the \gaia DR4 data reduction \citep{2021A&A...649A..11R}, might be able to produce a more accurate description of how the flux of bright and faint point-like sources is actually distributed on the detector.

\begin{figure*}
    \centering
    \includegraphics[width=0.97\textwidth]{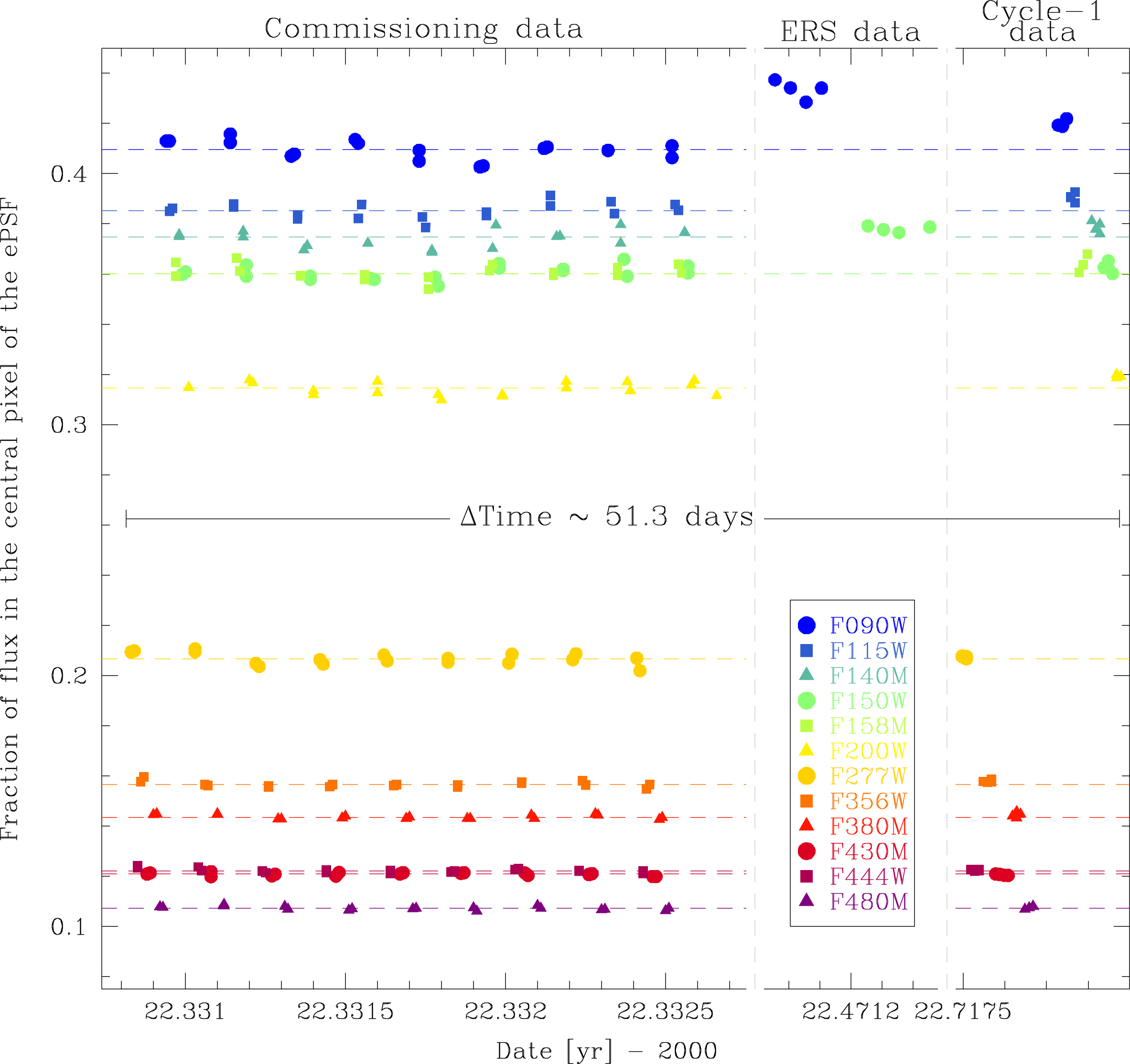}
    \caption{Temporal variation of the fraction of flux in the central pixel of the ePSF. Here we show the variations for the centermost ePSFs of the 5$\times$5 array. These ePSFs were obtained by perturbing the library ePSF for each image in the Commissioning, ERS and Cycle-1 data (see the text for details). The Commissioning data refer to all points before the date marked by the leftmost gray, dashed, vertical line. Points corresponding to the ERS data are those between the two gray, dashed, vertical lines. Finally, the Cycle-1 data comprise all remaining points in the plot. The dashed, horizontal lines color-coded as in the legend are the median values of the fraction of flux in the central pixel for the Commissioning ePSFs.}
    \label{fig:peak_psf}
\end{figure*}

\subsection{Temporal stability}\label{sec:psftime}

The true PSF of \jwst is expected to change over time, primarily due to small thermal variations, mirror movements including from occasional structural relaxations (``tilt events''), and mirror corrective moves from time to time. We monitored the temporal stability of our library ePSF models in two ways.

First, we studied the variation of flux in the centermost pixel of the library ePSFs (see Fig.~\ref{fig:peak_psf} for variations at the center of the NIRISS detector) using all data sets at our disposal, which cover a temporal baseline of $\sim$51 days. The library ePSF models were perturbed on an image-by-image basis following the prescriptions given in, e.g., \citet{2017ApJ...842....6B} and \citet{2018ApJ...861...99L,2022ApJ...934..150L}. The ePSF perturbation is an iterative process consisting of increasingly adjusting the models using residuals of the fit of well-measured sources. We report in the figure these perturbed central pixel values as a function of time for all 12 filters. The P2SF of \jwst appears to be relatively stable over the available temporal baseline, with deviations below $\sim$5\% except for the F090W and F150W ERS data, for which we measured a central pixel variation of about $\sim$7\% and 6\%, respectively. This appears to be consistent with the stability of the Optical Telescope Element of \jwst discussed in \citet{2023arXiv230101779M}. Temporal variations appear to be less prominent in redder filters.

\begin{figure}
    \centering
    \includegraphics[width=\columnwidth]{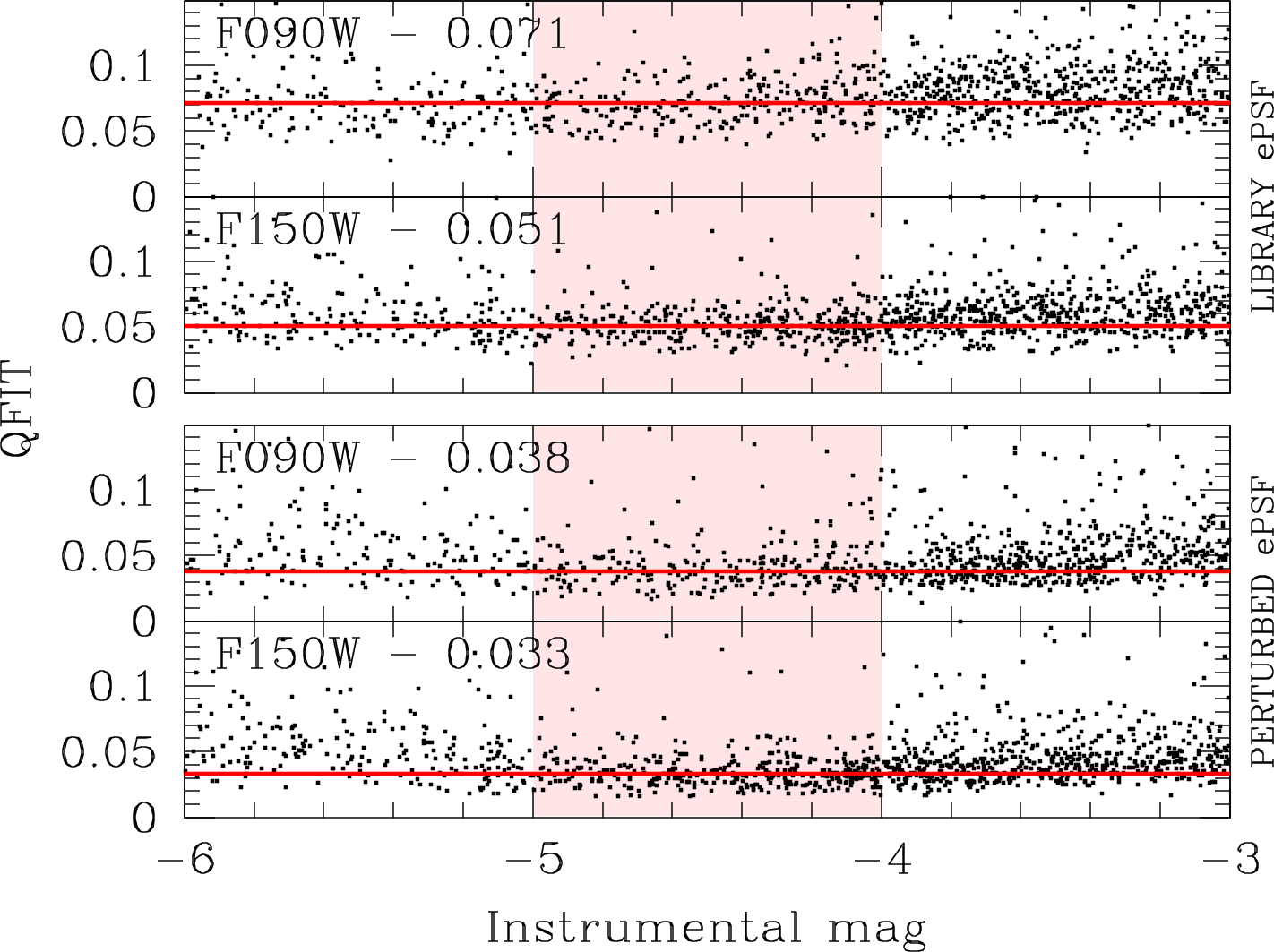}
    \caption{\qfit as a function of instrumental magnitude for two ERS images taken with the F090W and F150W filters. The top two panels are the results of using the Commissioning-based library ePSFs, while the bottom two panels refer to perturbed ePSFs.}
    \label{fig:m92_qfit}
\end{figure}

The second test consisted of comparing the \qfit of well-measured, bright stars obtained by fitting the library ePSFs and that coming from perturbed models. We narrowed the investigation to just the ERS data where the largest discrepancies were found in the previous test. The top two panels of Fig.~\ref{fig:m92_qfit} show the results using the library ePSFs, while the bottom two panels are for perturbed ePSFs. It is clear that the perturbed models provide a better representation of point-like sources (lower \qfit values) in the ERS data, pointing to a real change in the telescope PSF during this time.

It is clear from these tests that perturbing the library ePSFs to better match the PSF in an individual exposure can result in significantly better fits to stars. Since the perturbation process does not include information from multiple dithers, it should not be expected to significantly improve astrometry, but it can improve photometry (particularly if the perturbation is not constant across the detector) and can significantly improve star-galaxy discrimination and multiple-star fitting procedures. One must be careful, however, to ensure a minimum number of high signal-to-noise stars in each perturbation zone.

\section{Geometric-distortion correction}\label{sec:gd}

In this section, we describe the steps that lead us to achieve a very precise GD correction. In general, there are two main ways to solve for the GD \citep[see an extensive discussion in, e.g.,][]{2019JATIS...5d4005W}. The self-calibration approach makes use of observations characterized by  large dither patterns and different roll angles. In this way, a star is imaged in different parts of the detector, ideally from corner to corner of the FoV. This helps to randomize the direction of systematic GD residuals, so that the average stellar position on a common reference frame provides a somewhat ``distortion-free" representation of its true position that can be used to solve for the GD \citep[e.g.][]{2006acs..rept....1A,2014A&A...563A..80L,2015MNRAS.450.1664L,2021MNRAS.503.1490H}. This approach works best when many observations at different dither positions are available, and hence requires a considerable amount of observing time, thus not a viable option for \jwst given that sufficiently diverse data are not available at this early phase of the mission.

The second approach, which is the one that was officially adopted by the \jwst instrument teams during Commissioning, calibrates the GD through an external catalog.\footnote{A combination of these two approaches will likely be used for the \textit{Nancy Grace Roman Space Telescope} \citep{2019JATIS...5d4005W}.} To this end, \citet{2021jwst.rept.7716A} used two \hst data sets of the same LMC field taken 11 years apart to construct a high-precision, astro-photometric catalog for the GD correction of \jwst's imagers. This reference catalog offers the ability to calibrate \jwst's astrometry to a precision level similar to that of \hst.

We also decided to follow the second approach based on the \citet{2021jwst.rept.7716A} catalog as a reference, with the goal of achieving an astrometric precision better than 0.01 pixel. Unaccounted-for internal motions of LMC stars ($\lesssim$50 km s$^{-1}$, which at a distance of 49.5 kpc, correspond to 2.1 mas or $\sim$0.03 NIRISS pixel over 10 years) would prevent us from reliably detecting and correcting fine details in the GD of NIRISS. To mitigate this issue, we epoch-matched stellar positions of \citeauthor{2021jwst.rept.7716A}'s catalog to the average epoch of the NIRISS Commissioning observations using the PM information provided by the catalog. We additionally selected only stars that are bright and relatively isolated in NIRISS images. 

We started by measuring stellar positions in each NIRISS exposure by fitting the corresponding perturbed ePSF models, and kept only those with \qfit lower than 0.1 for the next stages. In the following, we describe each step of our GD correction using the F277W-filter data as guidance, but the same methodology and conclusions extend to all filters.

\subsection{Polynomial correction}\label{sec:poly}

\begin{figure*}
    \centering
    \includegraphics[width=\columnwidth]{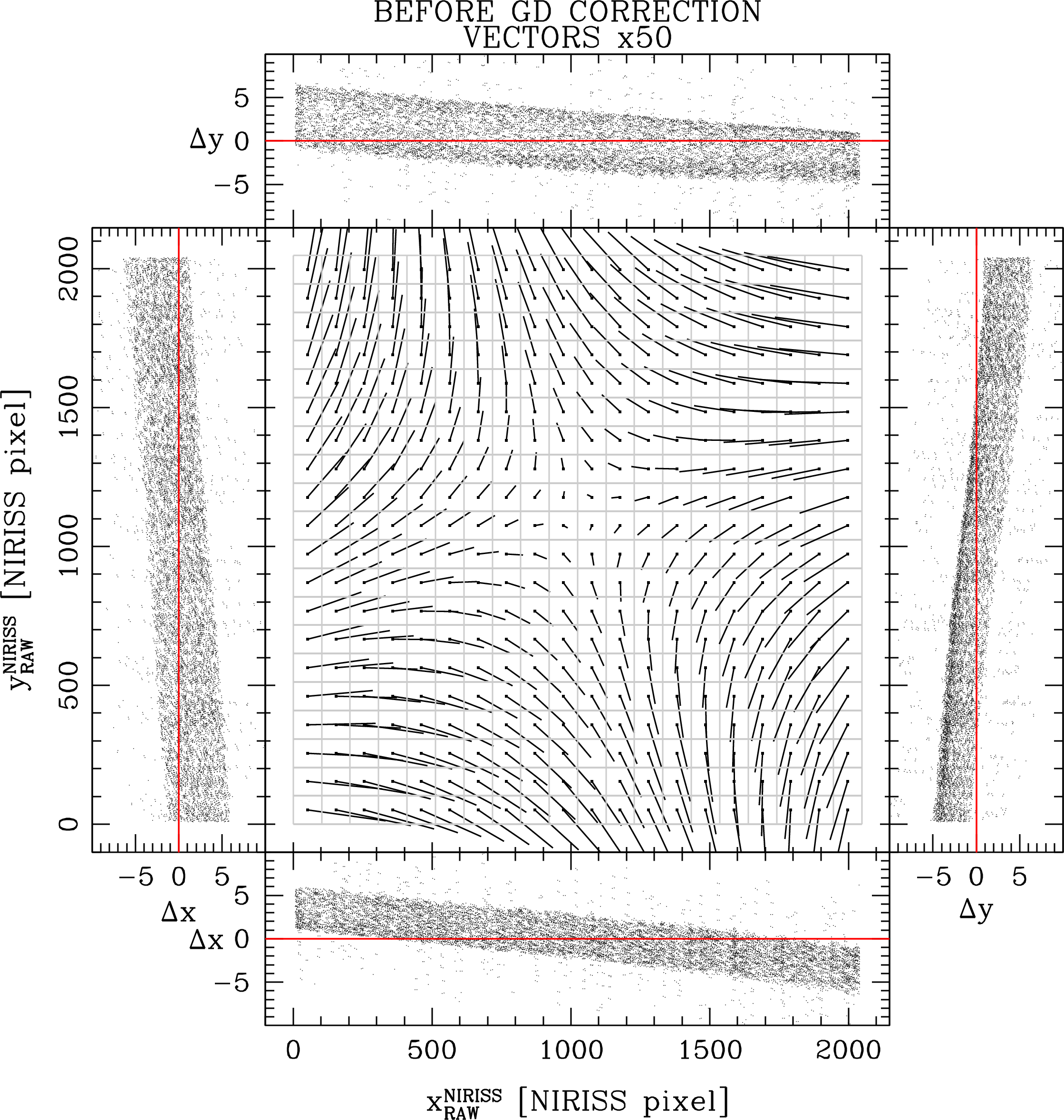}
    \hspace{0.5 cm}
    \includegraphics[width=\columnwidth]{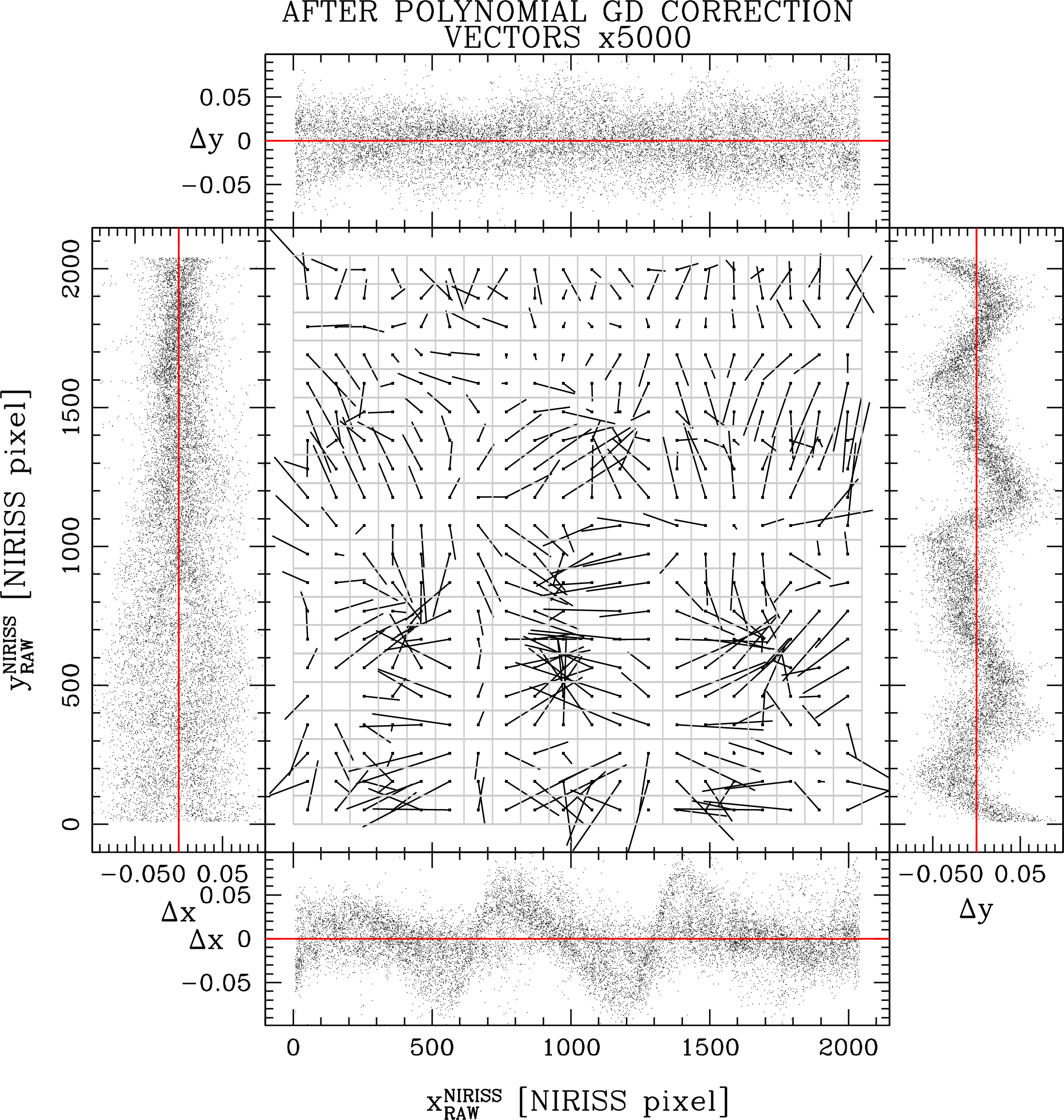}
    \caption{F277W-filter, GD-residual maps before (left) and after (right) applying the polynomial correction. The size of the vectors in the central square plots is magnified by factors of 50 (left) and 5\,000 (right) to show detail. The side panels show the $x$ and $y$ positional residuals (in units of NIRISS pixels) as a function of $x$ and $y$ raw NIRISS coordinates. In general, residuals improved by a factor $\sim$100 with the application of the polynomial correction.}
    \label{fig:gdpoly}
\end{figure*}

The bulk of the non-linear terms of the GD were fit with two fifth-order polynomial functions relative to the center of the NIRISS detector: $(x_{\textrm{ref}},y_{\textrm{ref}}) = (1024,1024)$ pixels. In addition, the use of normalized positions:
\begin{equation*}
    \left\{
    \begin{array}{c} 
        \Tilde{x} = \frac{x-x_{\textrm{ref}}}{x_{\textrm{ref}}} \\
        \Tilde{y} = \frac{y-y_{\textrm{ref}}}{y_{\textrm{ref}}}
    \end{array}
    \right . 
\end{equation*}
guarantees that the values of the coefficients of each polynomial term tell us the size of the distortion, center-to-corner, of that term in pixel units. The distortion correction for each star is defined as:
\begin{equation*}
\left\{
\begin{split}
    \delta x = \sum_{i = 1, 5}\sum_{j = 1, 5-i} a_{ij} \Tilde{x}^i\Tilde{y}^j \\
    \delta y = \sum_{i = 1, 5}\sum_{j = 1, 5-i} b_{ij} \Tilde{x}^i\Tilde{y}^j
\end{split}
\right . 
\end{equation*}

The values of the $a_{ij}$ and $b_{ij}$ coefficients were determined through an iterative procedure as described in \citet{2010A&A...517A..34B}. First, we cross-identified stars between each NIRISS catalog and the \hst reference catalog, and transformed  the epoch-matched \hst positions of common stars on to the raw coordinate frame of each NIRISS catalog by means of four-parameter linear transformations (two offsets, one rotation and one change of scale). The use of four parameters here instead of the usual six allows us to include the off-axis linear ``skew'' terms of the GD in the polynomial solution.

Once in the same raw coordinate frame, each cross-identification generates a pair of positional residuals $(\Delta x,\Delta y)$ defined as the difference between the transformed \hst positions and the raw NIRISS positions. These residuals were iteratively 3$\sigma$-clipped and averaged into a look-up table with 20$\times$20 square elements covering the full NIRISS detector, to minimize single-measurement noise. The averaged residuals were least-squares fit to obtain the coefficients of the two fifth-order polynomial functions. To ensure a smooth convergence of the solution, only half the values of the computed coefficients were added to the tally, and the process was repeated until the values in the look-up tables were smaller than $10^{-5}$ pixel.

Figure~\ref{fig:gdpoly} shows the F277W, two-dimensional distortion maps and the ($\Delta x$,$\Delta y$) positional residuals as a function of the $x$ and $y$ raw NIRISS coordinates before (left panels) and after (right panels) our polynomial correction. The NIRISS distortion has a magnitude of about 10 pixels from corner to corner of the FoV, with an RMS of 2.6 pixels.  After the polynomial correction, the impact of systematic residuals is significantly lower, reaching a corner-to-corner amplitude of $\sim$0.1 NIRISS pixel and an RMS of 0.03 pixel. To show detail, the size of the vectors in the two-dimensional maps is magnified by factors of 50 and 5\,000 for the before and after correction, respectively. Note that the orientation of the two-dimensional residuals after the correction is applied is significantly not random; this is even more evident by looking at the wavy residual patterns in the $\Delta x$ versus $x$ and $\Delta y$ versus $y$ side panels on the right.

\subsection{Look-up table of residual corrections}\label{sec:table}

\begin{figure}
    \centering
    \includegraphics[width=\columnwidth]{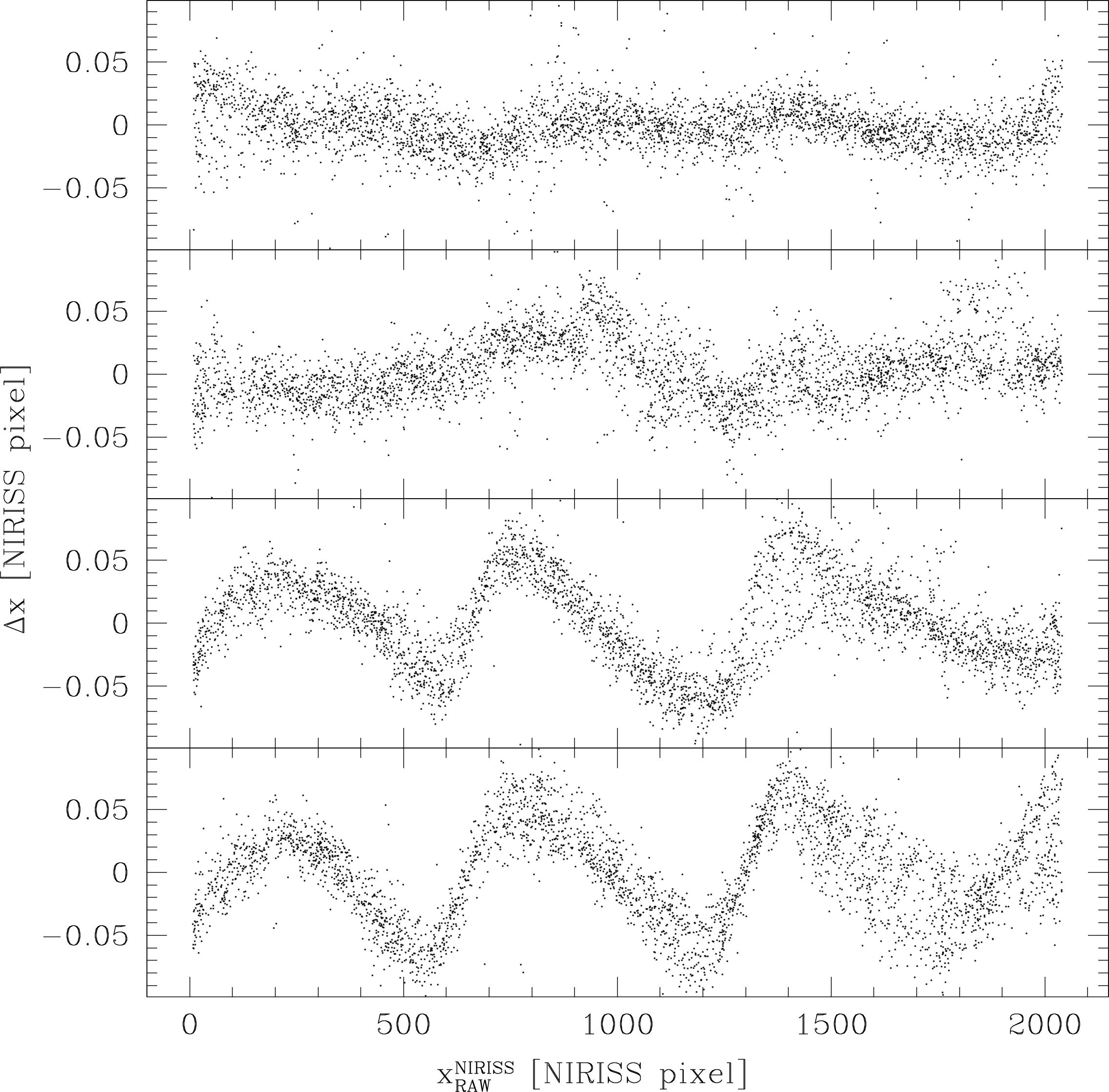}
    \caption{$\Delta x$ residuals as a function of $x$ raw NIRISS positions after applying the polynomial correction. Again, this plot refers to the F277W Commissioning data. Each panel shows the residuals for stars in four different stripes of 512 pixels each along the $y$ axis.}
    \label{fig:gdpolyres}
\end{figure}

It is clear from Fig.~\ref{fig:gdpoly} that there are higher-frequency variations in the distortion that cannot be accounted for with our fifth-order polynomial corrections. A better view of the wavy pattern present in the side panels of Fig.~\ref{fig:gdpoly} is provided in Fig.~\ref{fig:gdpolyres}, where $\Delta x$ residuals are shown as a function of $x$ position in four different 512-pixel-tall bands. The NIRISS readout electronics consist of four amplifiers that read 512 rows of pixels each, suggesting readout electronics could, at least in part, be responsible for the presence of these wavy patterns, in a similar fashion to the hysteresis effects in the readout electronics of the \mbox{HAWK-I@VLT} camera \citep[see, e.g.,][]{2014A&A...563A..80L}.

\begin{figure}
    \centering
    \includegraphics[width=\columnwidth]{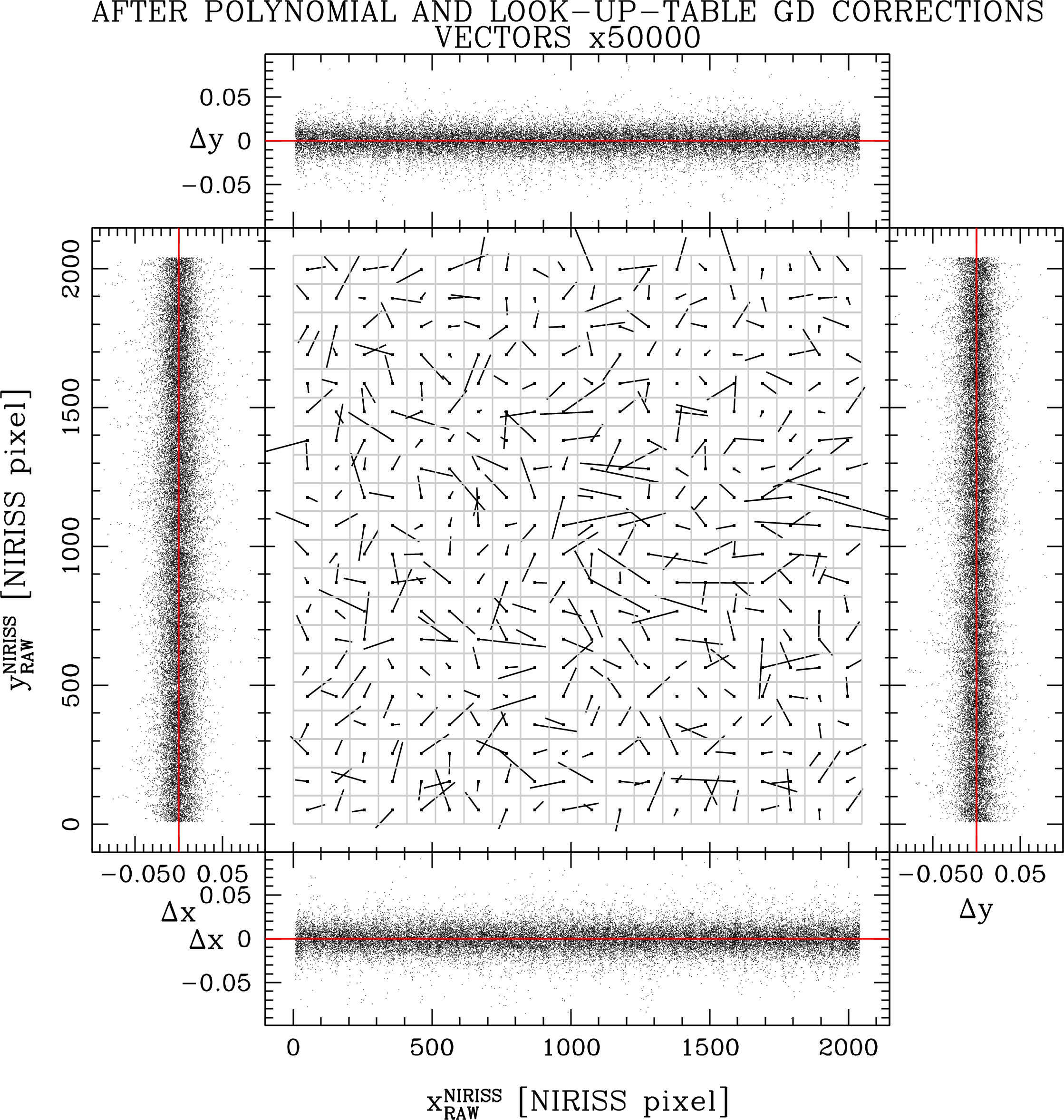}
    \caption{Similar to Fig.~\ref{fig:gdpoly}, but also including the {look-up} table correction. The size of the residual vectors in the two-dimensional map is now magnified by a factor 50\,000. No clear systematic trends are left.}
    \label{fig:gdcorr}
\end{figure}

Regardless of the true nature of these patterns --the investigation of which is outside the scope of the present work-- their correction can be empirically modeled using a look-up table of residuals, as done in, e.g., \citet{2011PASP..123..622B}. Polynomial-correction residuals were collected and averaged (with iterative 3$\sigma$ clipping) over a 32$\times$32 look-up table (square elements of 64 pixels per side), which allowed us to remove the vast majority of what was left after the polynomial correction. The look-up table is populated in such a way that elements adjacent to the edges of the detectors are applied as if they were measured exactly on the edges, to allow the use of bi-linear interpolation of the correction across the entire table \citep[see the scheme in][]{2014A&A...563A..80L}. As for the polynomial-correction derivation, we iterated the look-up table part of the distortion solution and only applied 50\% of the correction values each time until convergence, which was reached after 250 iterations.

\begin{figure}[t!]
    \centering
    \includegraphics[width=\columnwidth]{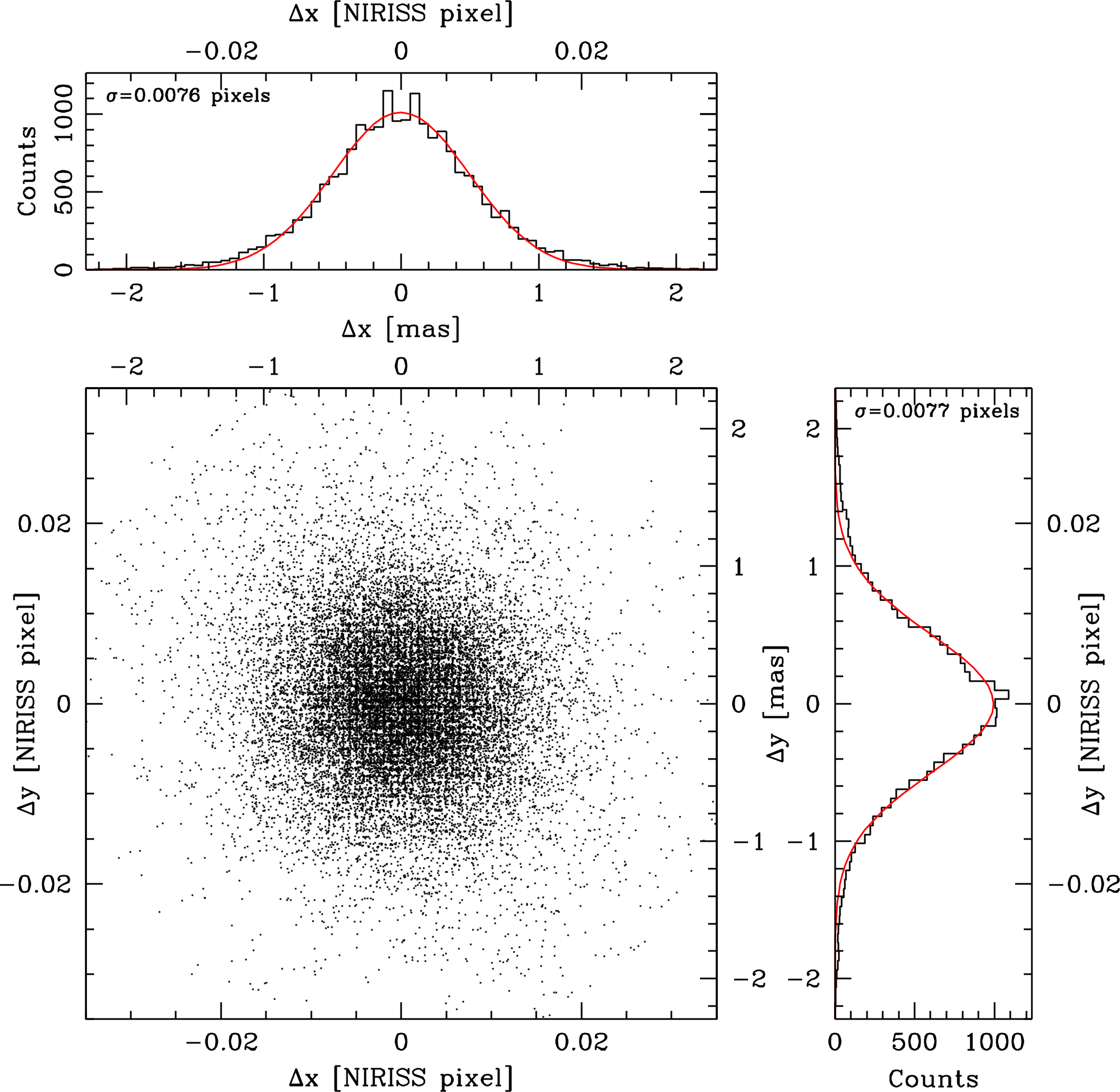}
    \caption{$\Delta y$ versus $\Delta x$ positional residuals of all 18 F277W single-exposure catalogs transformed onto the master frame. Here we only show residuals for well-measured stars in the instrumental-magnitude range between $-6.5$ and $-5$. Axis units are provided in both NIRISS pixels and mas. The one-dimensional histograms on the top and right-hand sides are fit with Gaussian functions (red curve) and the derived $\sigma$ are reported. The (1) ``roundness'' of the two-dimensional scatter plot, (2) $\sigma$ of the fit Gaussian curves, and (3) apparent lack of a statistically-significant deviation from Gaussianity in the wings of the histogram distributions are all \emph{clear} indicators of the high quality of the derived GD solution. Similar results are obtained for the other filters.}
    \label{fig:mastermatres}
\end{figure}

The two-dimensional distortion map and the one-dimensional positional residuals after both the polynomial and the look-up table residual corrections were applied are shown in Fig.~\ref{fig:gdcorr}. The figure shows no significant residual trends in either the $x$ or $y$ direction. The RMS of the positional residuals is $\lesssim$0.01 NIRISS pixel.

\begin{figure*}[t!]
\centering
\includegraphics[width=\textwidth]{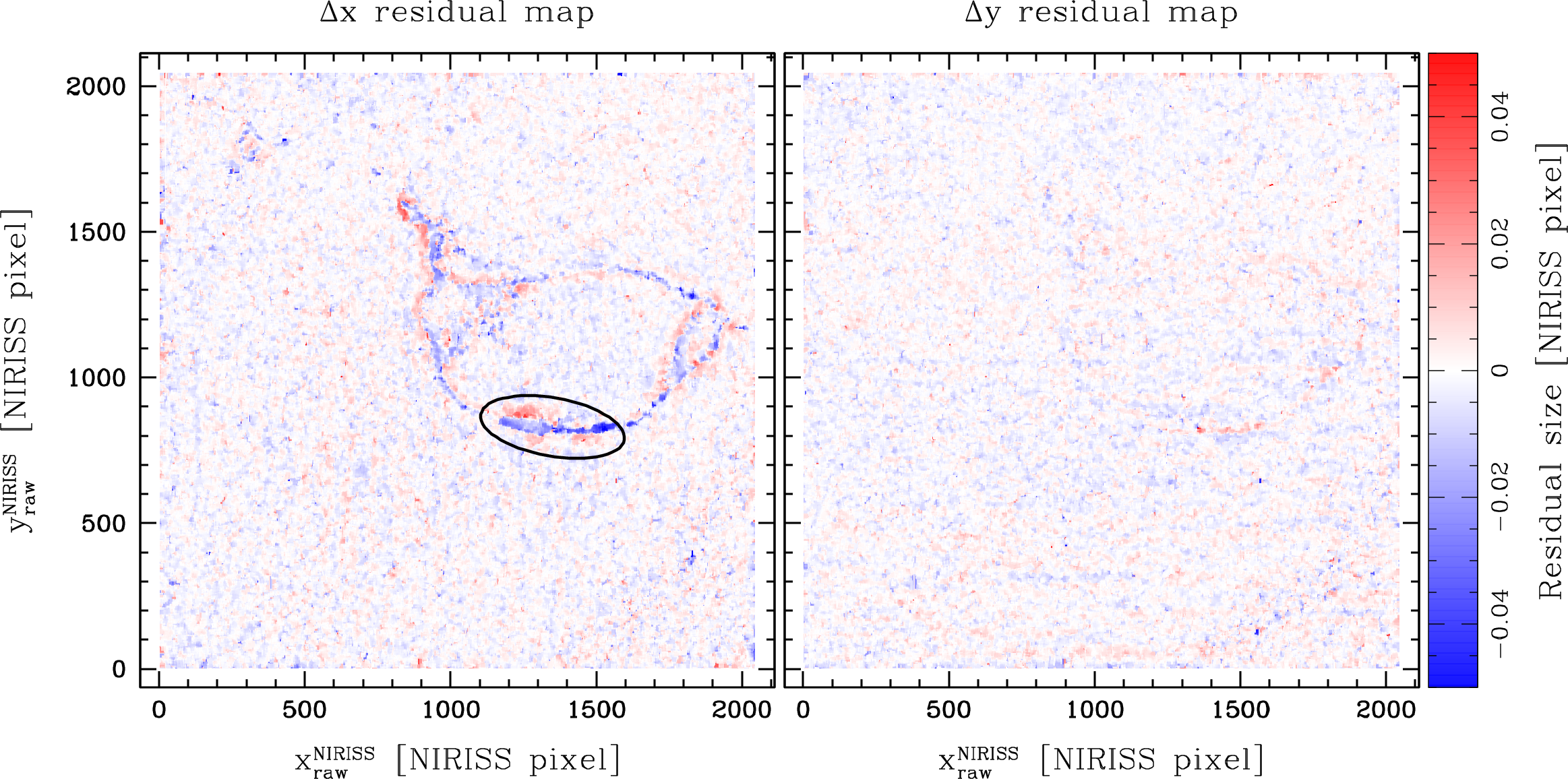}
\caption{Two-dimensional residual maps made by combining data from all filters (about 1\,000\,000 individual measurements of bright stars). These maps minimize filter-dependent effects, which are expected to be independent from one another, and enhance detector/electronics-related effects. The map on the left is for the $x$ direction and clearly shows a systematic pattern that is also visible, albeit less clearly, in the right map for the $y$ direction. Also note the presence of vertical (left panel) and horizontal (right panel) banding. The black ellipse in the left panel indicates the location of the suspicious feature identified when scanning the residuals.}
\label{fig:epoxyresmap}
\end{figure*}

Another quantitative assessment of the goodness of our distortion solution was obtained as follows. We constructed a common master frame of averaged positions by combining all 18 F277W single-exposure catalogs corrected with our polynomial and look-up table corrections. We then inverted the averaged positions of about 10\,000 well-measured stars with instrumental magnitude between $-6.5$ and $-5$ back into the raw coordinate systems of each exposure, and collected the differences $(\Delta x,\Delta y)$ between them and the corresponding single-exposure measurements. These residuals are shown in Fig.~\ref{fig:mastermatres}. We also show the one-dimensional histogram of the residuals along each axis and fit them with Gaussian functions (red lines in the side panels). It is clear that: (i) the two-dimensional distribution of the positional residuals is ``round'', which is what we would expect when no systematic errors are present; (ii) the $\sigma$ of the fitted Gaussian functions are of the order of 0.01 NIRISS pixel or less, which is comparable to the state-of-the-art distortion solutions of \hst's imagers; and (iii) the one-dimensional distributions are almost-perfectly fit by Gaussian functions, without any statistically-significant deviation in the wings. These are all indicators of the high quality of our GD correction.

\begin{figure}[t!]
    \centering
    \includegraphics[width=\columnwidth]{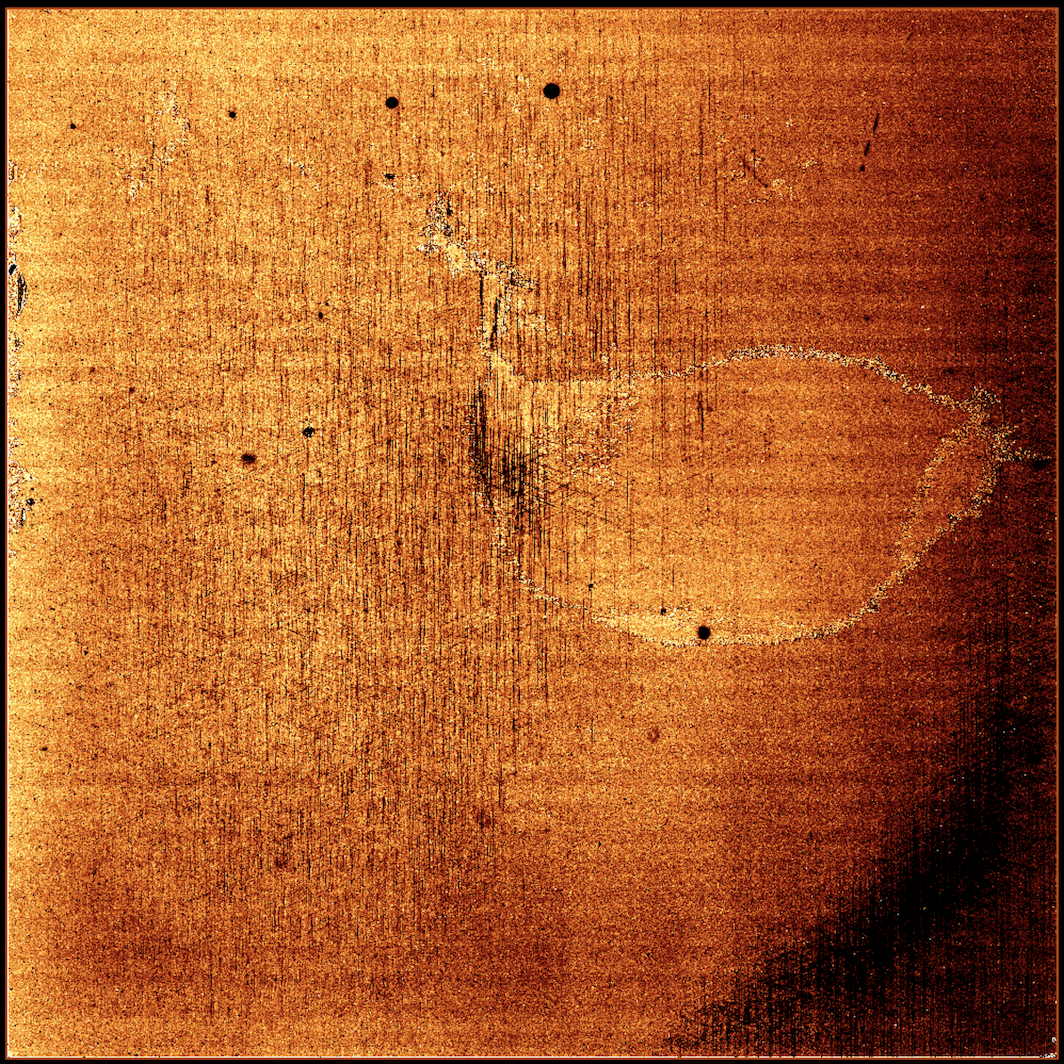}
    \caption{NIRISS F277W flat-field reference file from CRDS. Note the figure clearly resembles the pattern we found in the distortion residuals. There are also a few other smaller trends throughout the detector that correspond to secondary distortion features in the residual maps.}
    \label{fig:flat}
\end{figure}

\subsection{Epoxy voids}\label{sec:epoxy}

\begin{figure*}[t!]
\centering
\includegraphics[width=\textwidth]{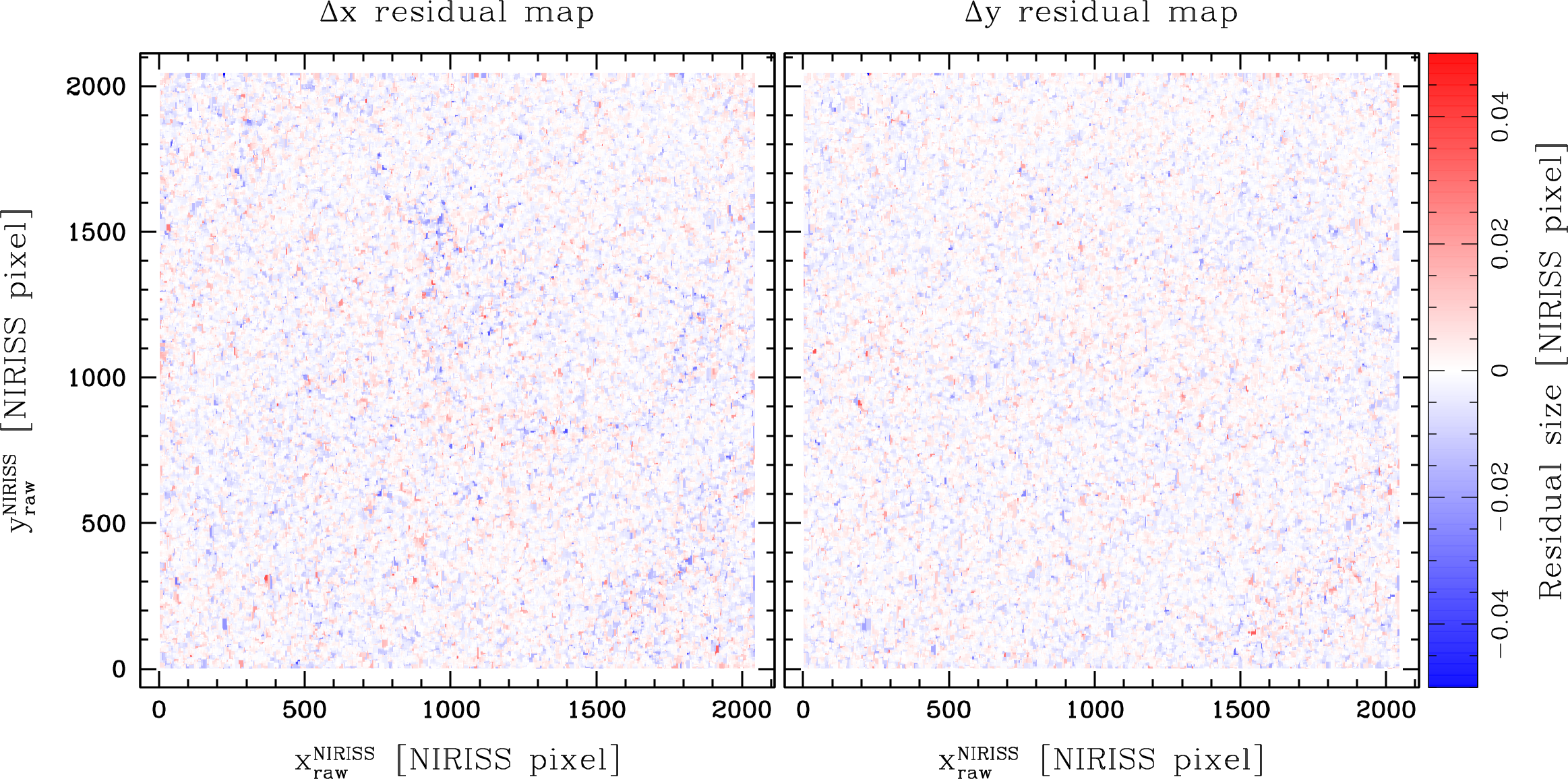}
\caption{Similar to Fig.~\ref{fig:epoxyresmap} but including the epoxy-related correction in the GD solution.}
\label{fig:epoxyresmapcorr}
\end{figure*}

The results shown in Figs.~\ref{fig:gdcorr} and \ref{fig:mastermatres} clearly demonstrate the adequacy of our GD solutions, and we could have stopped here without further tests and checks. Nonetheless, we decided to look at the F277W residuals one more time, scanning through the $x$ and $y$ raw-coordinate space in 10-pixel-wide bands. It is during this final test that we noticed a suspicious, possibly significant and localized $\Delta x$ residual trend for $y$ coordinates between 800 and 900 pixels. A negative $\Delta x$ peak, of about 0.05 pixel, appeared around $x = 1550$ pixels, and progressively shifted to $x = 1250$ while becoming positive by approximately the same amount as we progressed the $x$-coordinate scan from $y = 800$ to $y = 900$. We also noticed a few other possible smaller systematic trends in different locations on the detector, but their significance was at best marginal.

We then looked at the residuals in the other filters, and found very similar trends at around the same location in raw-coordinate space and with about the same amplitudes in all filters. These findings suggest that what we saw while scanning the F277W residuals could be a real distortion feature, rather than something related to ePSF models and/or a particular filter, and strongly favors a detector origin.

If indeed the origin is related to the detector, by collating the distortion residuals from all images of all filters we should be able to  maximize its signal. The two panels of Fig.~\ref{fig:epoxyresmap} show the results coming from this collation of $\sim$1\,000\,000 residuals. A distinct structure, mainly oval in shape and as large as 1/6 of the detector, stands out in the $\Delta x$ residual map on the left, and traces of it are also present in the $\Delta y$ residual map on the right. An ellipse on the left panel marks the location of the trend initially identified during the F277W scanning. Also worth noting is what seems to be a vertical-band pattern in the left panel, and a similarly-spaced, horizontal-band pattern in the right panel.

It is usually the case that pixels affected by detector-related residual patterns also stand out in flat-field images \citep[see, e.g.,][]{2011PASP..123..622B}. Figure~\ref{fig:flat} shows the F277W flat-field calibration reference file obtained from the \jwst CRDS. The large pattern identified in Fig.~\ref{fig:epoxyresmap}, as well as other smaller substructures related to secondary residual patterns are clearly visible.

\begin{figure*}[t!]
\centering
\includegraphics[width=\textwidth]{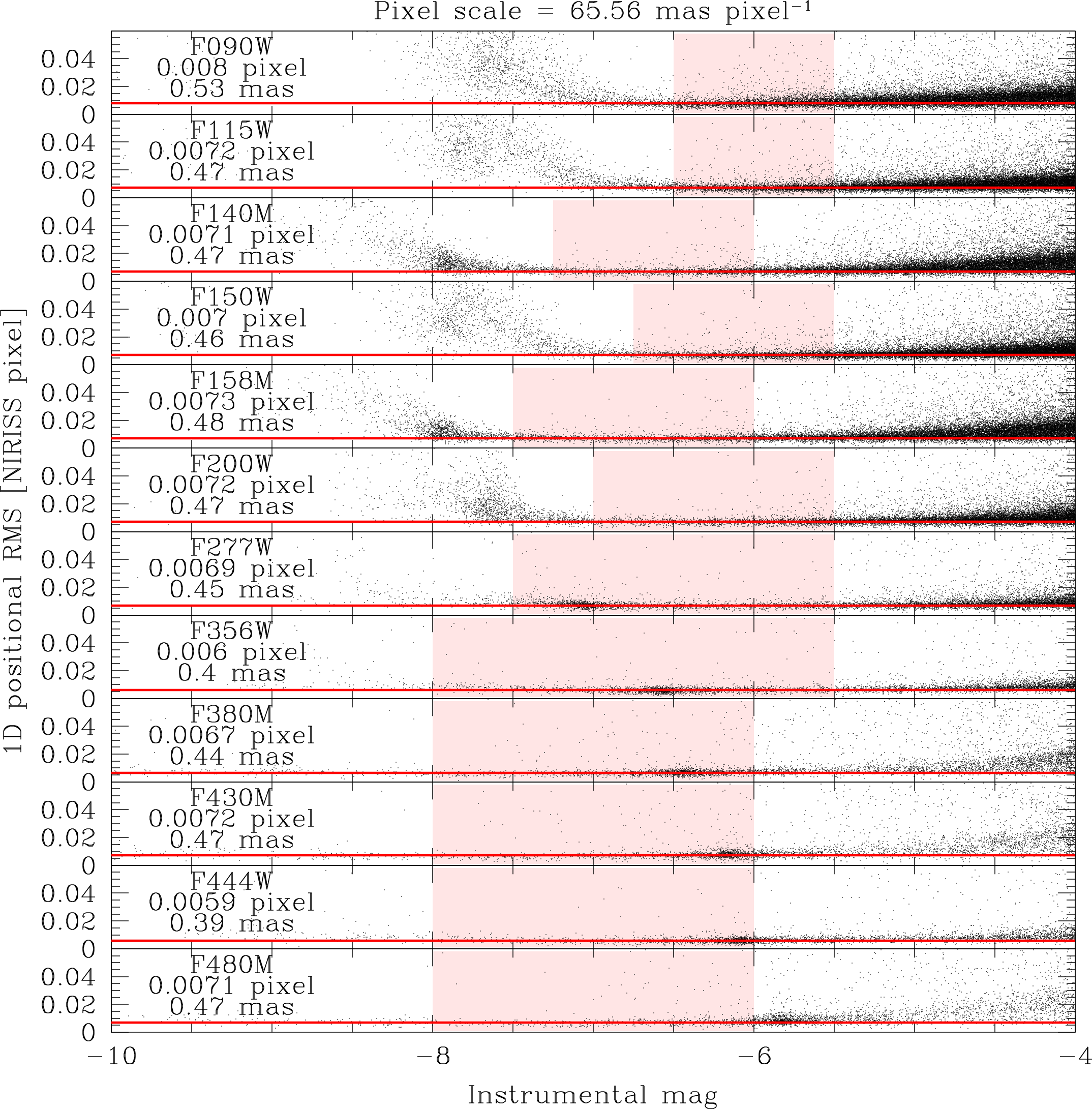}
\caption{One-dimensional positional RMS as a function of instrumental magnitude in all filters for the commissioning data. The red lines in each panel are set at the median values of the RMS for the best-measured stars that lie within the shaded pink regions. These median values are provided in each panel both in NIRISS pixels and mas (obtained using the average pixel scale listed at the top).}
\label{fig:rmspos}
\end{figure*}

Most mercury cadmium telluride (HgCdTe) diodes are grown on a substrate that is then removed to increase sensitivity at short wavelengths below 800 nm and to lower dark-current levels \citep[e.g.,][]{1221913}. To improve mechanical stability, the gap between the HgCdTe diode and the multiplexer is then filled with epoxy \citep[see also][]{2006PASP..118.1443B}. The features we identified in the distortion residuals and in the flat-field reference file are found in correspondence to voids in the epoxy underfill. While epoxy voids have been studied with respect to dark current and photometric performance\footnote{For the specific case of NIRISS, see: \href{https://jwst-docs.stsci.edu/jwst-near-infrared-imager-and-slitless-spectrograph/niriss-instrumentation/niriss-detector-overview/niriss-detector-performance}{https://jwst-docs.stsci.edu/jwst-near-infrared-imager-and-slitless-spectrograph/niriss-instrumentation/niriss-detector-overview/niriss-detector-performance}.} \citep[e.g.,][]{2020JATIS...6a6001R}, to our knowledge no one was expecting them to also have a measurable impact on astrometry, as we have discovered here for the NIRISS detector.

To account for these fine structures in our distortion solution, we opted for a pixel-based correction rather than an additional look-up table of residuals. The correction is made up of two parts, one for the $x$ direction and one for the $y$ direction, and is derived as follows: we assigned to each pixel of the 2048$\times$2048 pixel array the 2.5$\sigma$-clipped median of the closest 50 residuals coming from all filters. This was done as a compromise between smoothing out most of the noise coming from the individual measurements and preserving the residual patterns at both small and large scales. After including this new correction, the systematic trends in the positional residuals have been significantly reduced (Fig.~\ref{fig:epoxyresmapcorr}).

\begin{figure*}[t!]
\centering
\includegraphics[width=\textwidth]{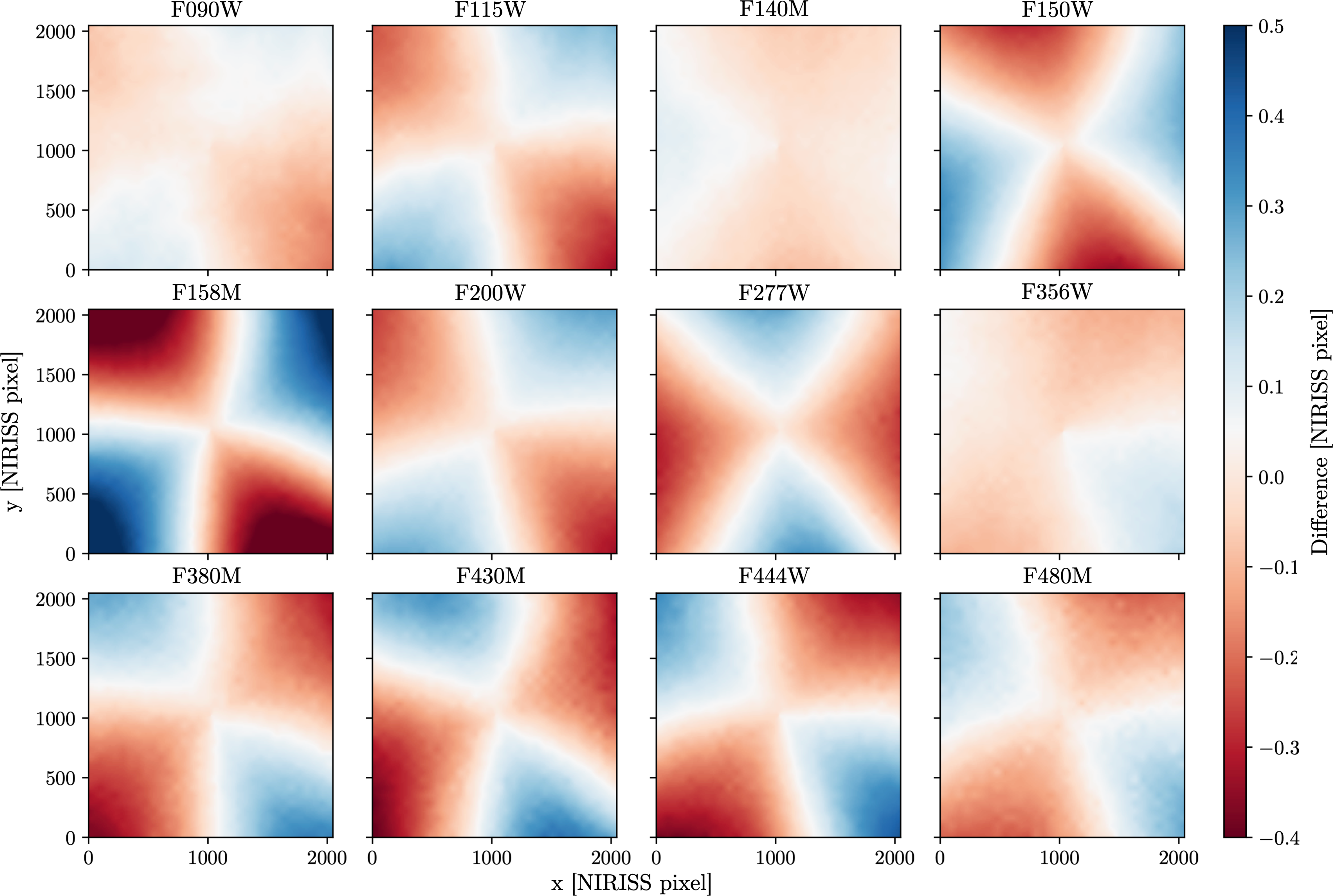}
\caption{Differences between the filter-averaged and the filter-dependent GD corrections in units of NIRISS pixels. All plots have the same linear color scale shown in the colorbar on the right. Filter-based residuals can be as large as 0.5 NIRISS pixel.}
\label{fig:filtvar}
\end{figure*}

\subsection{Astrometric precision}\label{sec:precision}

The distortion maps and residuals presented in the previous Sections are clear qualitative indicators of the accuracy of our GD corrections. Quantitatively, state-of-the-art GD corrections for \hst have residuals $\lesssim$0.01 pixel \citep[e.g.,][]{2006acs..rept....1A,2011PASP..123..622B}, measured as the RMS around the mean of averaged dithered positions for the same sources. Note that these residuals are the sum in quadrature of ePSF-fitting errors and the GD-correction errors themselves. Therefore, they are more an indicator of how precise the whole astrometry package is than just the GD correction alone.

Our quantitative assessment of the astrometric precision reached with the NIRISS Commissioning data was done as follows. For each filter, we averaged together positions of sources measured in all NIRISS single-exposure catalogs, using the full, 3-part GD solutions and six-parameter linear transformations. The process was iterated three times to improve both the averaged positions on the master frame and the transformations. Only stars measured in at least three images were considered in the analysis. The one-dimensional master-frame positional RMS as a function of instrumental magnitude in all filters are presented in Fig.~\ref{fig:rmspos}. The red lines in each panel are set at the median values of the RMS of the best-measured stars that lie within the shaded pink regions. Pixel values were transformed to mas using the average plate scale of all pixels, which is listed at the top of the Figure (see also Sect.~\ref{sec:scale}). For all filters, the median values of the positional RMS are well below 0.01 NIRISS pixel ($<$0.66 mas). Pixel-wise, the precision reached is superior to that achieved with \hst's imagers, which is expected given the more stable environment of \jwst.

\subsection{Filter dependence}\label{sec:gdfilt}

Any optical element traversed by light upstream of the detector can contribute to the GD. So, we investigated whether, and to what extent, the filter components affect the shape and amplitude of the GD. 

We averaged together the total two-dimensional GD correction of all filters to obtain a proxy for a filter-independent solution. Then, we subtracted it from the GD correction of each filter to isolate their filter-dependent component. The result is shown in Fig.~\ref{fig:filtvar}. The different patterns in each panel point to significant (up to 0.5-pixel) filter dependencies in the GD correction. Residuals $\lesssim$0.5 pixel are small enough for some scientific applications (e.g., catalog cross-matching), but are definitely too large for high-precision astrometric studies. This validates our choice of deriving filter-dependent GD corrections.~\\

\subsection{Pixel scale}\label{sec:scale}

The \hst catalog of \citet{2021jwst.rept.7716A} provides us with an opportunity to measure the absolute pixel scale of the NIRISS detector. For each filter, we cross-identified bright, well-measured stars between each single-exposure NIRISS catalog of the Commissioning data and the \hst catalog of \citet{2021jwst.rept.7716A} and used six-parameter linear transformations to relate the two sets of positions. One of these linear parameters is the relative scale between NIRISS and the \hst catalog, which was defined to be exactly 50 mas pixel$^{-1}$. We use this information to estimate the absolute pixel scale of NIRISS, allowing for a dependence on filter.

Velocity aberration causes cyclical changes in a detector's pixel scale as \jwst orbits the Earth-Sun system \citep[e.g., for \hst see][]{2003hstc.conf...58C}. Although small, these scale changes need to be accounted for in the computation of the absolute scale of a detector. The velocity-aberration correction factor is provided in the header of every NIRISS image with the keyword \texttt{VA\_SCALE}. We used it to correct the values of the NIRISS pixel scales that we derived from the \hst catalog cross-matching. It is worth mentioning that this correction is quite small for the specific observations used, because--by design--the LMC calibration field targeted for Commissioning is in the Continuous Viewing Zone of \jwst, and the pointing of the telescope there is almost always orthogonal to its orbit around the Sun.

\begin{figure*}[t!]
\centering
\includegraphics[width=\textwidth]{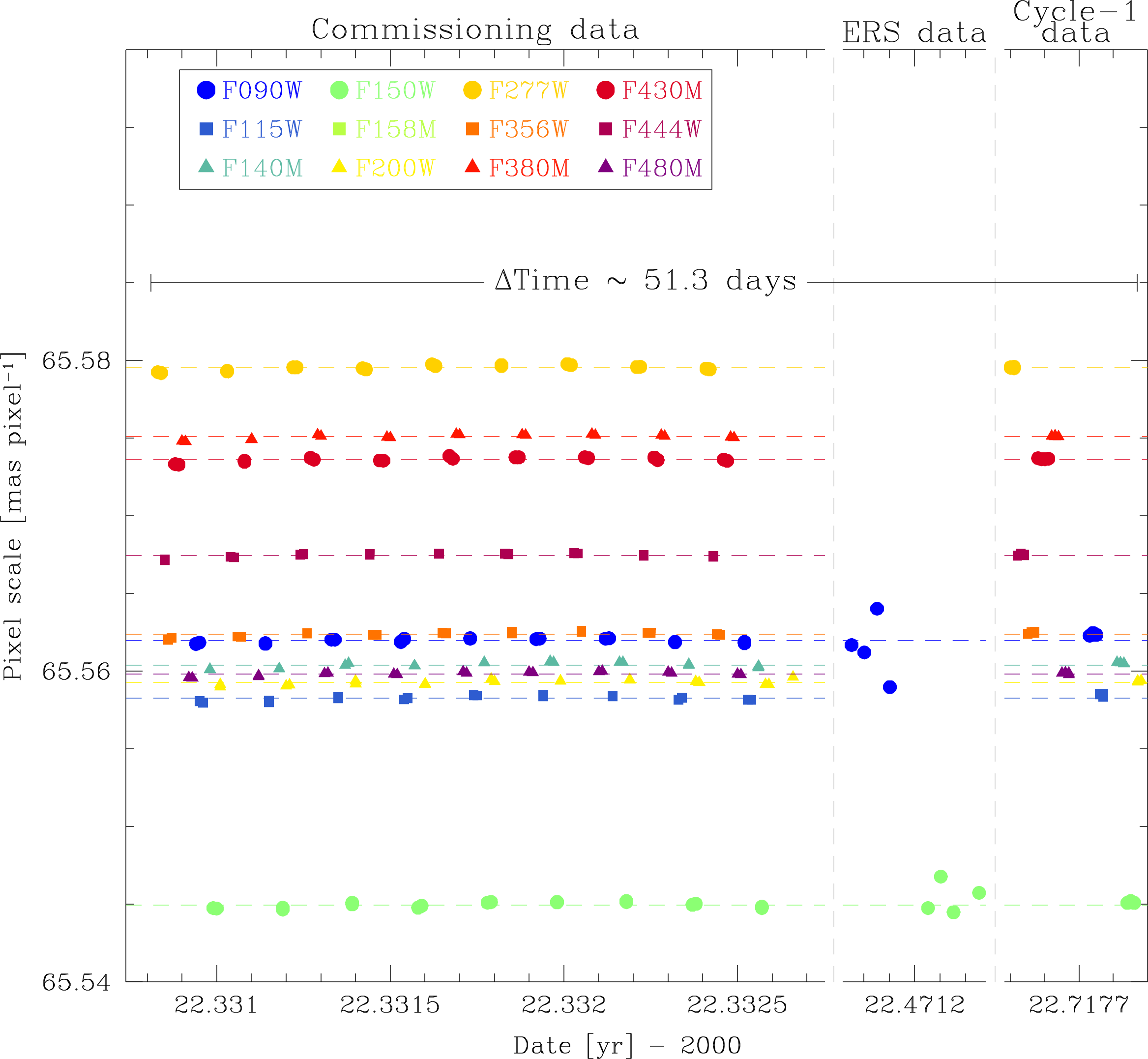}
\caption{Pixel scale as a function of time for the various filters.  All values are corrected for velocity aberration using the \texttt{VA\_SCALE} factor. The dashed, horizontal lines, color-coded as in the legend, are the average values of the pixel scale for the Commissioning data reported in Table~\ref{tab:scale}. The points related to the ERS data set present a scatter larger then the others because the cross-match with the \gaia catalog was only possible for faint \gaia stars (${\it G} > 19$ mag).}
\label{fig:scale}
\end{figure*}

Table~\ref{tab:scale} collects our measurements of the absolute pixel scales for each filter. We find significant scale differences among the filters, which is consistent with what is stated in the official documentation of NIRISS\footnote{\href{https://jwst-docs.stsci.edu/jwst-near-infrared-imager-and-slitless-spectrograph/niriss-instrumentation/niriss-detector-overview}{https://jwst-docs.stsci.edu/jwst-near-infrared-imager-and-slitless-spectrograph/niriss-instrumentation/niriss-detector-overview}.}. We measure the filter-averaged scale of NIRISS to be 65.562 $\pm$ 0.003 mas pixel$^{-1}$.

\begin{table}[t!]
    \centering
    \begin{tabular}{c|c|c}
        \hline
        \hline
        Filter & Pixel scale [mas pixel$^{-1}$] & Error [mas pixel$^{-1}$] \\
        \hline
        F090W & 65.56196 & 0.00003 \\
        F115W & 65.55825 & 0.00004 \\
        F140M & 65.56039 & 0.00004 \\
        F150W & 65.54495 & 0.00004 \\
        F158M & 65.53851 & 0.00003 \\
        F200W & 65.55927 & 0.00004 \\
        F277W & 65.57951 & 0.00004 \\
        F356W & 65.56238 & 0.00004 \\
        F380M & 65.57509 & 0.00003 \\
        F430M & 65.57361 & 0.00003 \\
        F444W & 65.56745 & 0.00003 \\
        F480M & 65.55983 & 0.00003 \\
    \hline 
    \hline
    \end{tabular}
    \caption{Average pixel scale of each filter corrected for velocity aberration. The filter-averaged scale of NIRISS that we measure is 65.562 $\pm$ 0.003 mas pixel$^{-1}$.}
    \label{tab:scale}
\end{table}

We also investigated the temporal variation of the pixel scale using all available data sets (Fig.~\ref{fig:scale}). For the ERS data, which are focused on the globular cluster M\,92, we used the \textit{Gaia} Data Release 3 (DR3) catalog \citep[projected on to a tangent plane centered at the center of one ERS NIRISS image, with a fixed scale of 66 mas pixel$^{-1}$]{2016A&A...595A...1G,2021A&A...649A...1G} as a reference to compute the pixel scale of each image. Unfortunately, the only stars in common between the ERS data and the \gaia catalog are at the faint end of \gaia (${\it G} > 19$), which resulted in less precise transformations in Fig.~\ref{fig:scale}.

We find the scale of Cycle-1 images to be significantly different (at the 8-$\sigma$ level) from that of Commissioning data for the F090W filter. The scale difference becomes progressively less noticeable the redder the filters, down to less than 0.7$\sigma$ for the F480M filter.

\begin{figure}[t!]
\centering
\includegraphics[width=\columnwidth]{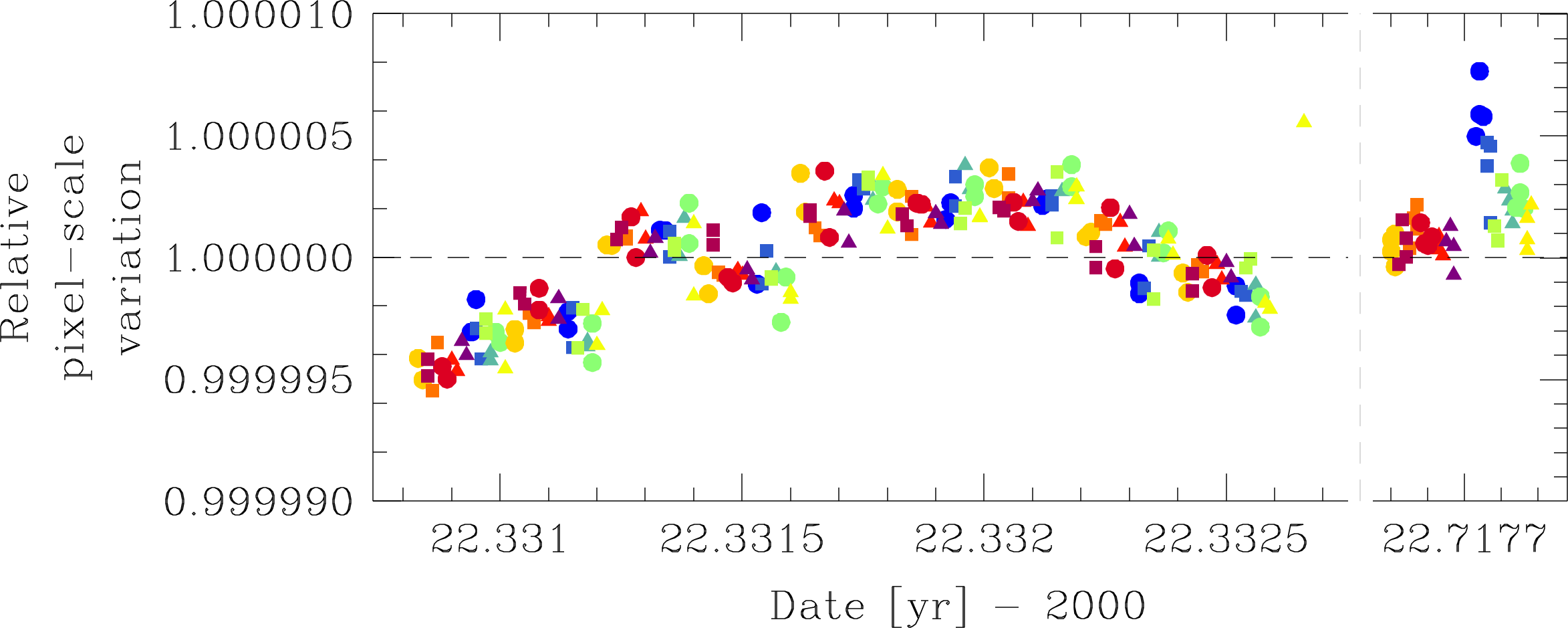}
\caption{Relative pixel-scale variation in Commissioning and Cycle-1 images. The shapes and colors of symbols are the same as in Fig.~\ref{fig:scale}. The black horizontal line is set at 1 for reference.}
\label{fig:scaleperc}
\end{figure}

Figure~\ref{fig:scaleperc} shows the relative pixel scale variation in Commissioning and Cycle 1 images. There is a clear apparent trend as a function of time during the $\sim$12-hour duration of the Commissioning observations. While many potentially contributing factors were changing during Commissioning, there were no commanded adjustments in telescope mirror alignment or instrument focus \citep{2023arXiv230101779M} over the time period of these measurements and, further, the observatory had already reached thermal stabilization weeks earlier, such that there were no substantial ongoing thermal drifts (Menzel et al. 2023, PASP in press). We also computed the scale against the \hst-based reference frame \citep{2021jwst.rept.7716A}, against a pure \gaia DR3 frame (for the reddest filter that has more stars with $G < 19$ in common with the \gaia catalog), and against the central-pointing NIRISS image, and in all cases saw a similar trend. If we hypothesize that the change in pixel scale is due to a drift in the Optical Telescope Element (OTE) secondary mirror position, the necessary amount of secondary-mirror defocus to change the pixel scale by $5 \times 10^{-6}$ according to optical modeling is significantly inconsistent with the measured stability in the wavefront sensing observations over this time period. The origin of these apparent changes in pixel scale remains unclear. Continued monitoring of the NIRISS pixel scale over time may help understand the cause of this subtle effect.

\section{Sample Scientific Application: Population-dependent kinematics of the LMC}\label{sec:science}

An immediate scientific application demonstrating the quality of our ePSF models and GD solutions is provided by the Commissioning data. We computed PMs of sources in the LMC calibration field by combining \hst and \jwst data. As a first-epoch data set, we took advantage of the catalog made by \citet{2021jwst.rept.7716A} using \hst data taken at epoch 2006.39. This catalog defines our first-epoch master frame. It is oriented with North up and East to the left, and it has scale of 50 mas pixel$^{-1}$. We created filter-dependent NIRISS master frames by averaging together positions and fluxes of common sources measured in the individual images. Orientation and scale for the NIRISS master frames were set up using the \hst catalog, while the photometry was registered to that of a NIRISS image at the center of the dither pattern. For PM purposes, we used the NIRISS F090W as our second-epoch master frame, since it provides the highest number of sources in common with the \hst catalog. The master frames of the remaining NIRISS filters are used for additional photometric information in our analysis.

\begin{figure}[t!]
\centering
\includegraphics[width=0.85\columnwidth]{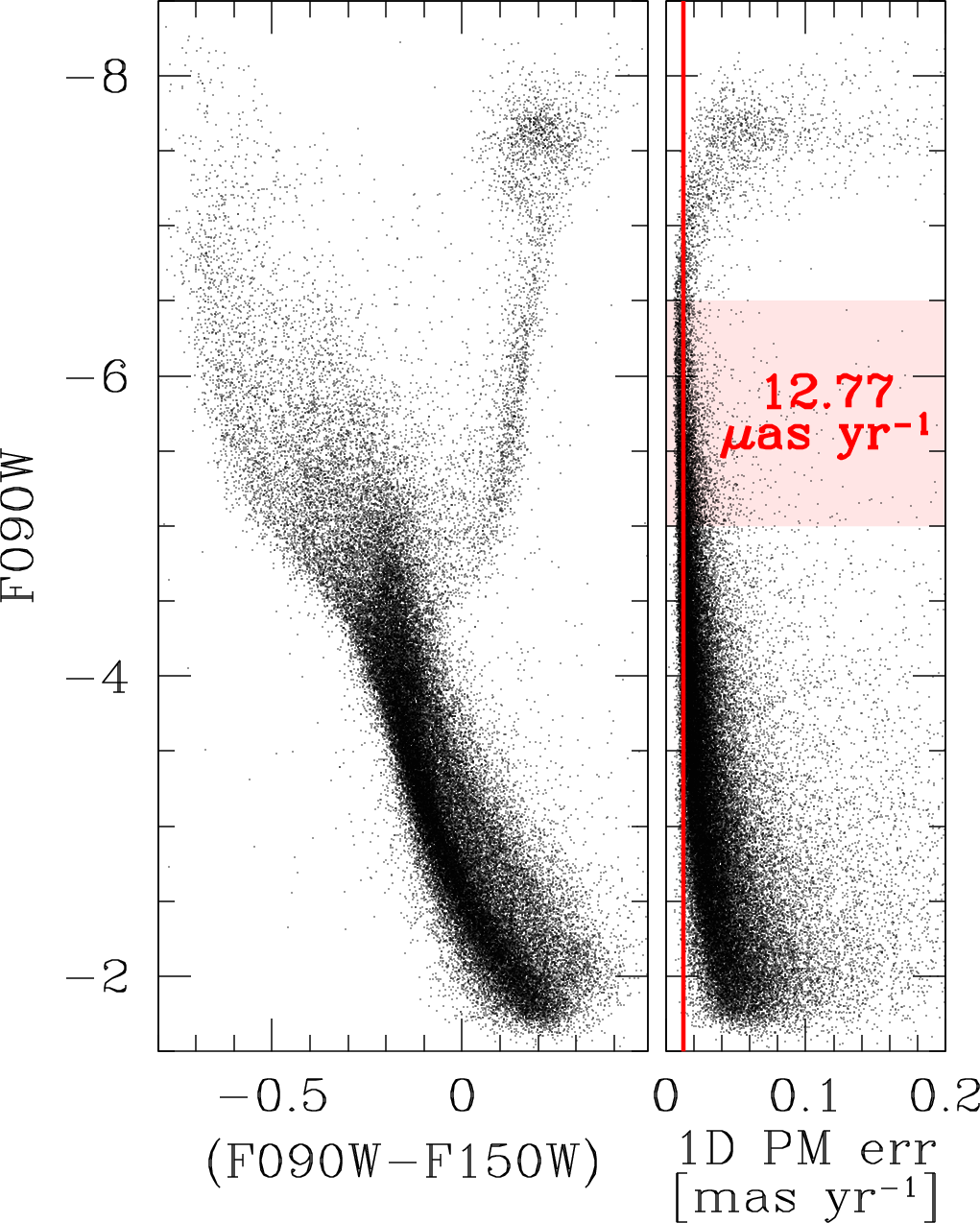}
\caption{The instrumental F090W versus (F090W$-$F150W) CMD of the stars in the LMC calibration field is shown in the left panel. The right panel presents the one-dimensional PM error as a function of instrumental F090W magnitude. The red vertical line is set at the median value (also reported in the plot) of the one-dimensional PM error for the best-measured stars within the shaded pink region.}
\label{fig:cmd090}

\centering
\includegraphics[width=0.99\columnwidth]{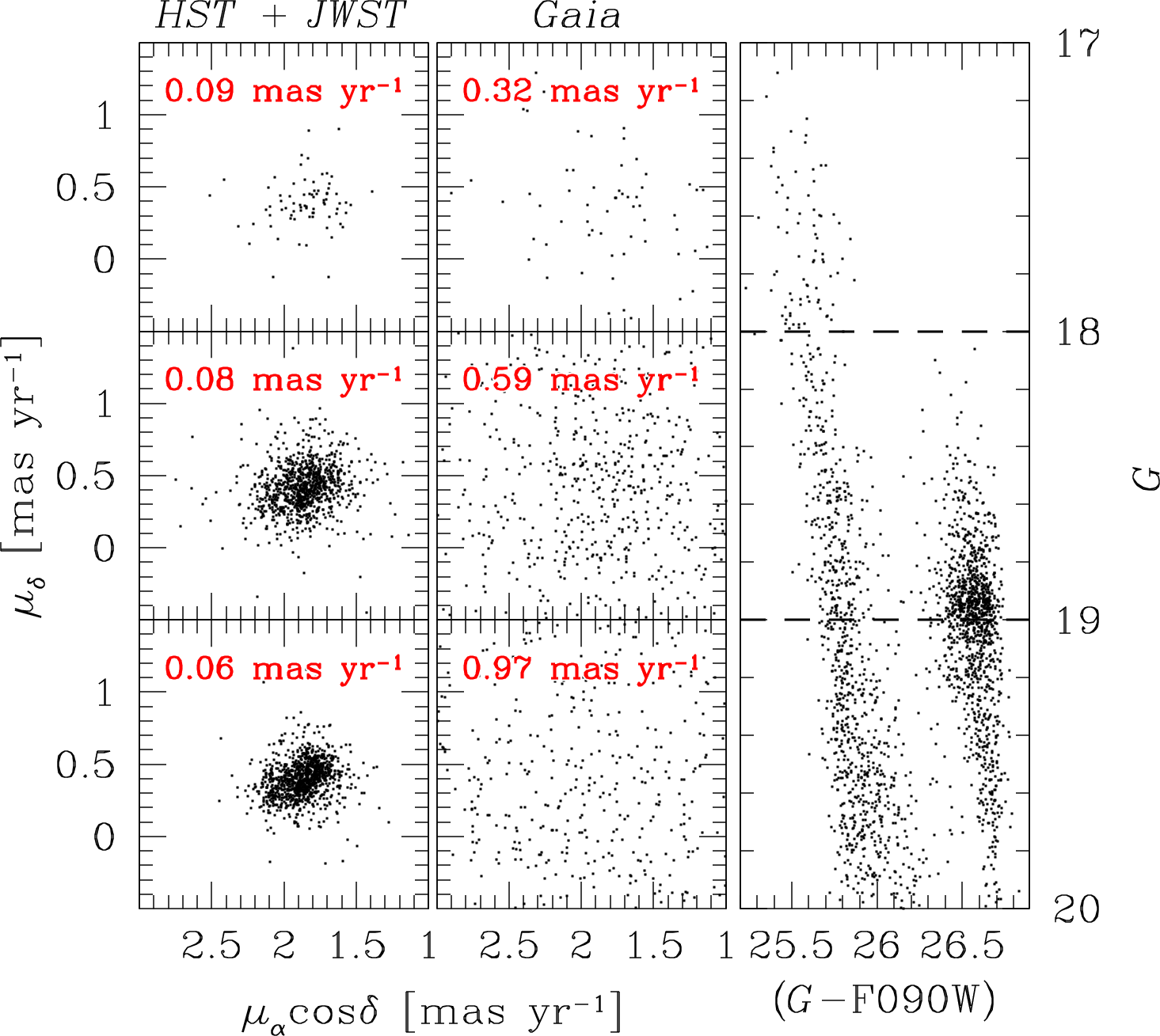}
\caption{Comparison between the \textit{HST}$+$\jwst (left panel) and the \gaia DR3 (middle panel) PMs. Each panel is divided into three parts pertaining to stars of different \textit{G} magnitude that are highlighted with dashed lines in the \textit{G} versus (\textit{G}$-$F090W) CMD in the right panel. The {\it observed} PM dispersion of stars is reported as a reference in each VPD.}
\label{fig:gaia}
\end{figure}

We transformed positions as measured in the NIRISS master frame into to the reference frame of the 2006.39 \hst catalog by means of general, six parameters linear transformations. The coefficients of these transformations were obtained using a set of bright, well-measured, old red-giant branch (RGB) stars --selected in the instrumental F090W versus (F090W$-$F150W) color-magnitude diagram (CMD)-- as reference stars, as shown in the left panel of Fig.~\ref{fig:cmd090}. RGB stars define a narrow sequence in the CMD, as opposed to the very broad blue sequence typical of rich centrally-located LMC fields, which is a superposition of populations with different ages, chemical compositions and, most importantly, kinematics \citep{2021A&A...649A...7G}.

Once in the same reference system, relative PMs were computed as the difference between the original \hst and transformed \jwst positions, divided by the temporal baseline ($\sim$15.94 yr). We multiplied the result by the first-epoch pixel scale of 50 mas pixel$^{-1}$ to obtain PMs in units of mas yr$^{-1}$. PM errors were estimated as the sum in quadrature of the positional errors of the two catalogs divided by the temporal baseline. Finally, our relative PMs were registered onto an absolute reference frame by means of the absolute PM zero-point provided by the comparison between our relative PMs and the absolute PMs in the  \citet{2021jwst.rept.7716A} catalog.

\begin{figure*}[t!]
\centering
\includegraphics[width=\textwidth]{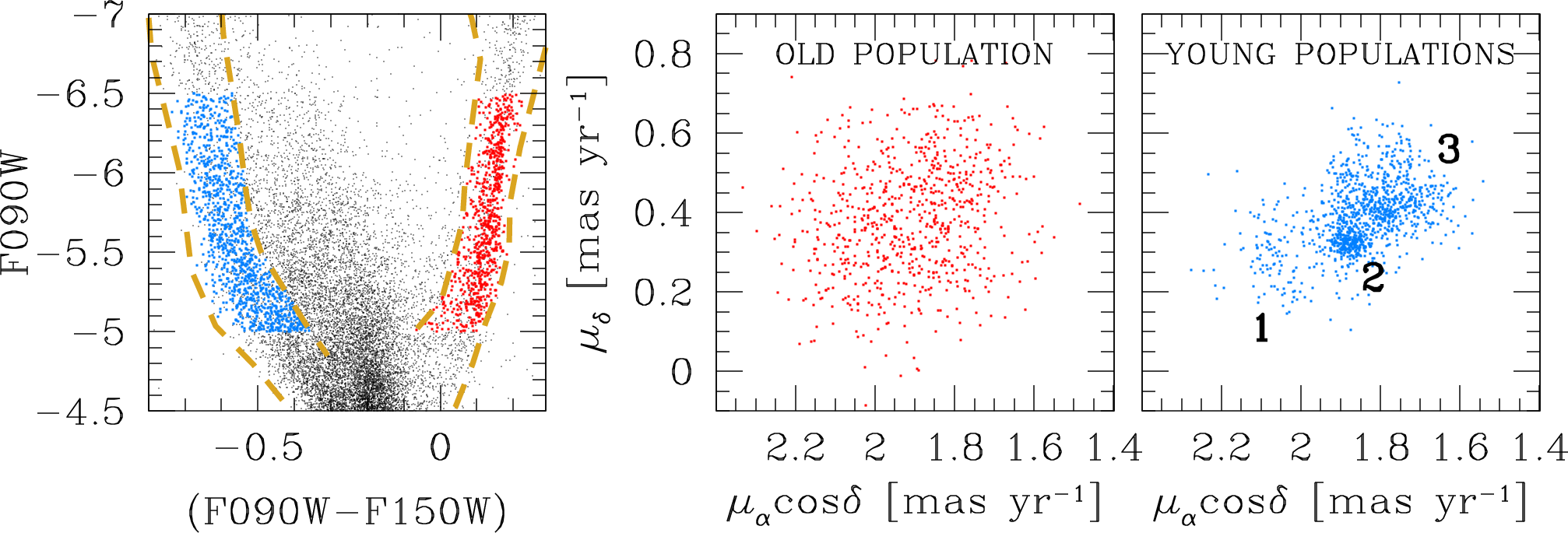}
\caption{The instrumental CMD in the left panel illustrates our selection procedures for the old (red) and the young (blue) populations. Stars are in the magnitude range $-6.5<\textrm{F090W}<-5$, where PM errors are the smallest. The dashed gold lines are used to constrain the two groups in the F090W$-$F150W color. The other two panels in the Figure present the PM distribution of the old and the young populations, color-coded as in the CMD. Three main substructures appear to be present in the rightmost VPD, and are labelled for clarity. The median PM uncertainty is $\sim 0.01$ mas yr$^{-1}$, cf.~Fig.~\ref{fig:cmd090}, well below the width of the (sub)structures visible in both VPDs.}
\label{fig:vpdoy}
\end{figure*}

The right panel in Fig.~\ref{fig:cmd090} presents the one-dimensional PM error as a function of F090W instrumental magnitude. For bright, well-measured stars in the instrumental magnitude range between $-6.5$ and $-5$, the median PM error is $\sim$13 $\mu$as yr$^{-1}$. This corresponds to 3.1 km s$^{-1}$ at the distance of the LMC \citep[1 mas yr$^{-1}$ $\sim$ 235 km s$^{-1}$;][]{2002AJ....124.2639V}. This is exquisite for studies of the internal kinematics of the LMC disk. Stars at these instrumental magnitudes have apparent magnitudes around ${\it G} \sim 20$ (given the ${\it G}-{\rm F090W}$ colors in the right panel of Fig.~\ref{fig:gaia}). By contrast, even for its best-measured stars ${\it G} \sim 9$--12, {\gaia} DR3 only achieves uncertainties of $\sim 16$--23 $\mu$as yr$^{-1}$ \citep[Table~4 of][]{2021A&A...649A...2L}. Around the LMC tip of the RGB with ${\it G} \sim 16$ \citep{2021A&A...649A...7G}, {\gaia} DR3 median errors are $\sim 38$--65 $\mu$as yr$^{-1}$, some 3--5 times larger than what we report here for stars that are some 4 magnitudes fainter. The uncertainties reported here are comparable to those obtained in state-of-the-art PM studies combining {\hst}-only data over a similar temporal baseline \citep[e.g.,][]{2014ApJ...797..115B,2018ApJ...854...45L,2022ApJ...934..150L}. This is a testament of the overall quality of our \jwst NIRISS ePSF and GD models.

\begin{figure}[t!]
\centering
\includegraphics[width=\columnwidth]{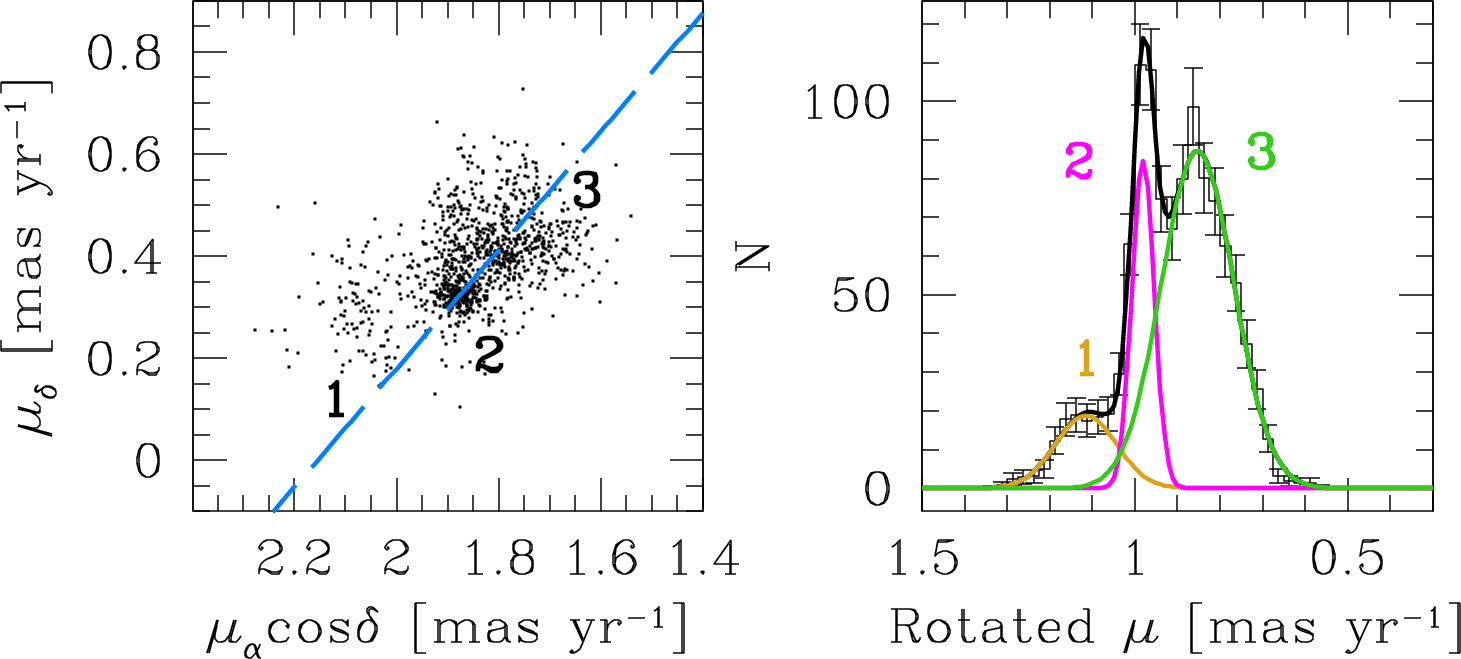}
\caption{The VPD of the young populations selected in Fig.~\ref{fig:vpdoy} is shown in the left panel. The blue dashed line passes through the elongation defined by Groups 2 and 3, and is used to construct the histogram in the right panel, with error bars. A triple Gaussian function is fitted to the histogram (black), with the three individual components colored in gold, magenta and green, from left to right, for Groups 1 to 3, respectively. The magenta component likely refers to stars in the cluster OGLE-CL LMC 407. In both panels, the three populations are labelled as in Fig.~\ref{fig:vpdoy}.}
\label{fig:rotvpd}
\end{figure}

We compared our PMs with those in the \gaia DR3 catalog. Unsaturated stars in common with the \gaia catalog are between ${\it G} = 17$ and  ${\it G} = 20$.  Figure~\ref{fig:gaia} summarizes the results of the comparison. The vector-point diagrams (VPDs) of stars in three magnitude bins -- each width ${\it G} = 1$~mag -- are shown for our \textit{HST}$+$\jwst PMs (left panels) and \gaia's PMs (middle panels). We report the observed PM dispersion in red in each panel. While \gaia's PM errors are larger than the intrinsic dispersion of stars in the field, our PMs are clearly able to reveal substructures in the internal motion of LMC stars.

These substructures are population dependent, as can be seen from our samples illustrated in Fig.~\ref{fig:vpdoy}. We selected the bluest and the reddest stars in the F090W versus (F090W$-$F150W) CMD (left panel in Fig.~\ref{fig:vpdoy}) within the instrumental magnitude range $-6.5<\rm{F090W}<-5$. This is the range where stars have the best PM precision according to Fig.~\ref{fig:cmd090}. Stars in the red selection are mostly old RGB stars (hereafter, the old population), while those in the blue selection are mostly a mix of young and intermediate-age sources with ages between 50 Myr and 2 Gyr \citep[hereafter, the ``young populations", e.g.,][]{2021A&A...649A...7G}.

\begin{figure*}[t!]
\centering
\includegraphics[width=\textwidth]{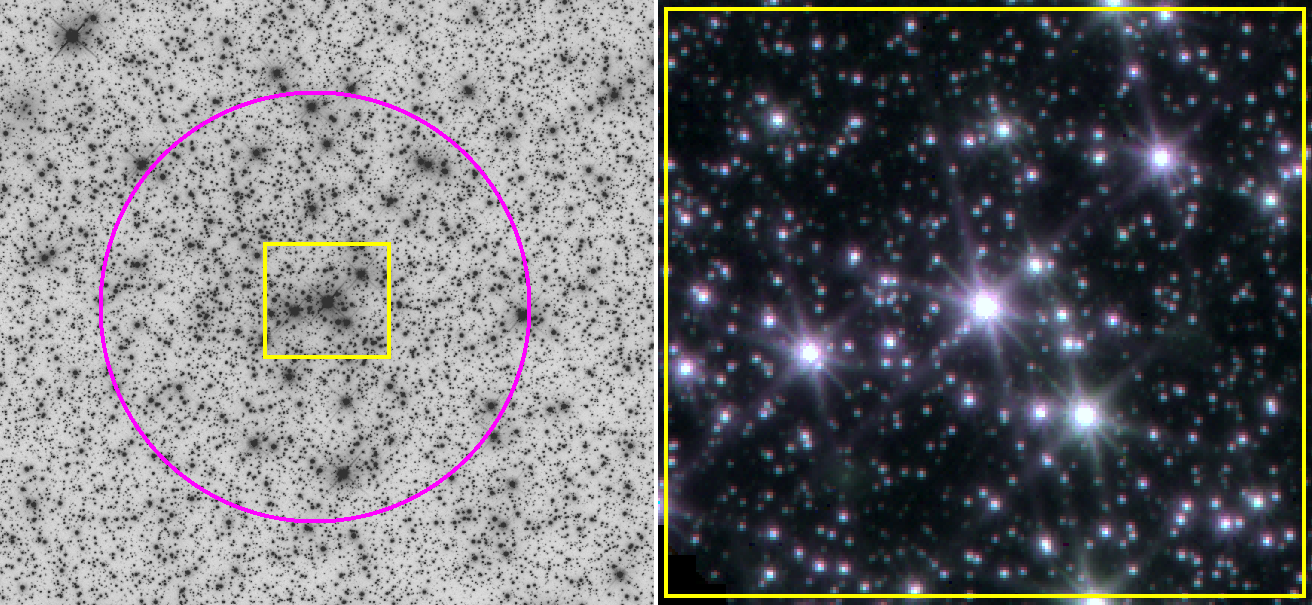}
\caption{The grey-scale stacked image made with data taken with the \hst Wide Field Channel (WFC) of the Advanced Camera for Surveys (ACS) is shown in the left panel. The magenta circle is centered on the overdensity of stars associated with OGLE-CL LMC 407, has a radius of 25 arcsec, and is used to refine our cluster-member selection (see the text for details). The region within the yellow rectangle of size 7.3$\times$6.6 arcsec$^2$ is zoomed-in and shown as seen by NIRISS in the right panel, where we present a trichromatic view made using the following color-scheme: blue $=$ F090W, green $=$ F115W, red $=$ F140M. In both panels, North is up and East is to the left.}
\label{fig:stack}
\end{figure*}

The distribution of stars in these two groups in the VPD is clearly different. The old population is characterized by a broad distribution centered around ($\mu_\alpha \cos \delta,\mu_\delta) = (1.90,0.38)$ mas yr$^{-1}$, whereas the young populations seem to be made up of three main groups (highlighted in the rightmost VPD in Fig.~\ref{fig:vpdoy}), with hints of some of them being further fragmented into additional substructures. Group 1 is the least populated and most isolated of the three; Group 2 seems to be the most concentrated, and it overlaps substantially with a broader Group 3.

Perhaps a clearer way to highlight substructures in the motion of the young populations is through the analysis of their histogram distribution. To this end, we used the methodology described in \citet{2017ApJ...844..164B} and summarized in Fig.~\ref{fig:rotvpd}. The distributions of the Group 2 and 3 stars in the VPD appear elongated along a specific direction (blue dashed line in the left panel). We projected the motions onto this line and then constructed a histogram (right panel). To remove any dependence on the bin width and the starting point of the histogram, we followed the approach of \citet{2019ApJ...873..109L} and used 1\,000 individual samplings of the data, each with a different amount of added PM noise according to the stars' PM-error distribution. We simultaneously fit three Gaussian functions to the resulting histogram that are shown in gold, magenta and green in the right panel of the figure, for Groups 1, 2 and 3, respectively. The areas of these Gaussians are indicative of the relative abundance of young stars of the three groups in our FoV. We find that about 12\% of the stars belong to Group 1, 20\% to Group 2 and the remaining 68\% to Group 3.

Among the young population substructures in the VPD, Group 2 stands out with the tightest distribution. It has  a dispersion of $\sim$0.026 mas yr$^{-1}$ (i.e., 6.2 km/s), measured as the $\sigma$ of the Gaussian fitted in Fig.~\ref{fig:rotvpd}. and it is centered at $(\mu_\alpha \cos\delta,\mu_\delta) \sim (1.8,0.3)$ mas yr$^{-1}$. Stars within $\pm$3$\sigma$ from the center of the magenta Gaussian are preferentially located in proximity of the South-East edge of the FoV, at odds with members of the other two groups that have a more uniform spatial distribution. 

Interestingly, Group-2 stars are found to be in the same region where the 120-Myr-old stellar cluster OGLE-CL LMC 407 is located \citep{2022A&A...666A..80N}, and are likely members of this cluster. We then further restricted Group-2 objects to only those within a 25$^{\prime\prime}$ radius from the center of the overdensity of stars (defined by hand) seemingly associated with the OGLE-CL LMC 407 cluster, which is visible in the stacked \hst  image from the epoch-one data (see Fig.~\ref{fig:stack}). Using this refined sample, we were able to measure the absolute PM of the cluster as the 3$\sigma$-clipped median of the distribution in the VPD:
\begin{equation*}
    (\mu_\alpha \cos\delta,\mu_\delta) = (1.873 \pm 0.003, 0.321 \pm 0.002) \textrm{ mas yr}^{-1}\, .
\end{equation*}

\begin{figure}[t!]
\centering
\includegraphics[width=\columnwidth]{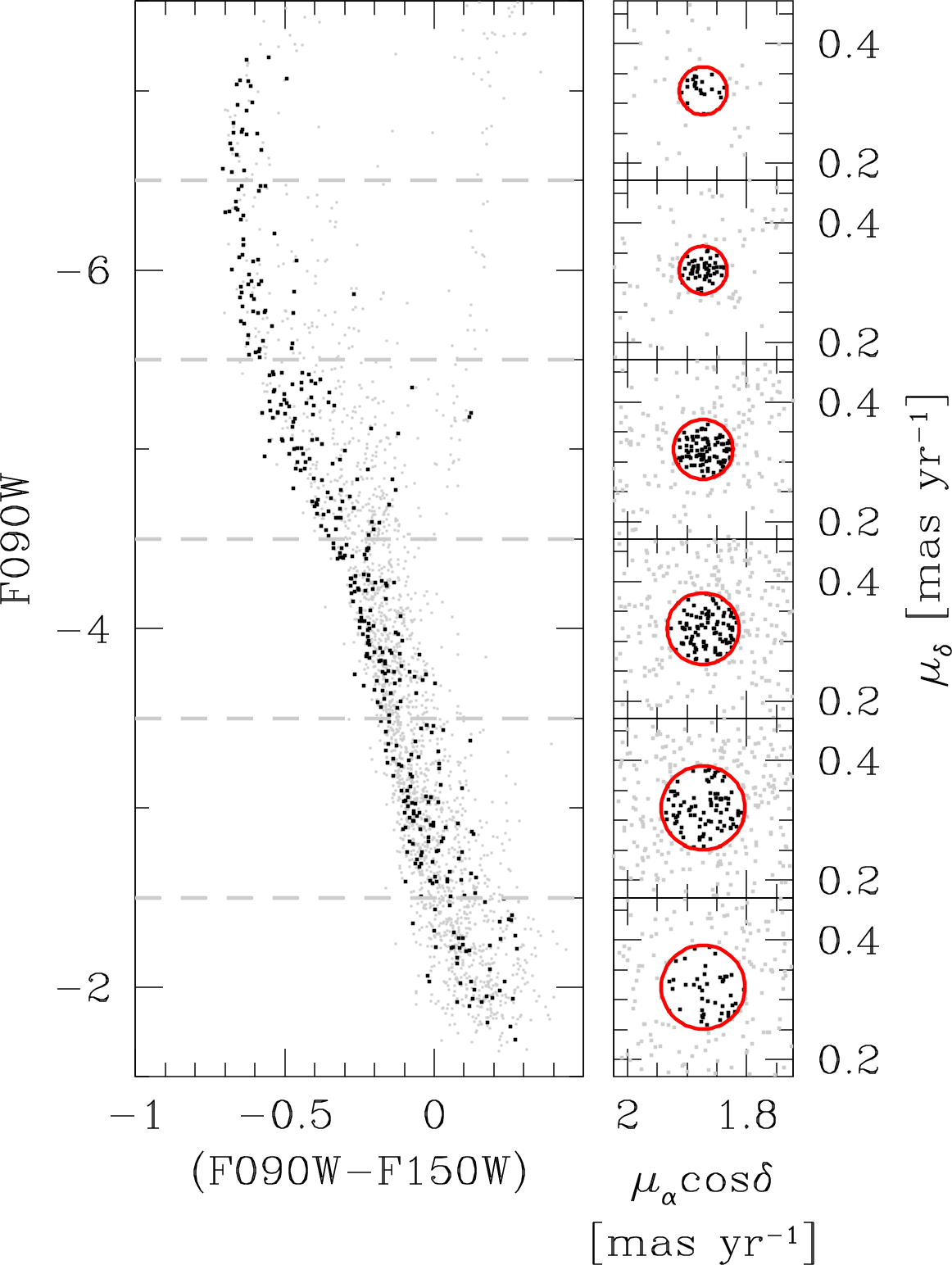}
\caption{Instrumental F090W versus (F090W$-$F150W) CMD of bona-fide OGLE-CL LMC 407 members within a radius of 25 arcsec from the cluster's center, and with PMs within the magnitude-dependent circles shown in red in the VPDs of the right panels (see the text for details). Cluster members are shown as black points, while all other stars are represented by grey dots. The radius of the circles ranges from 0.04 to 0.07 mas yr$^{-1}$, from the bright to the faint magnitude bin.}
\label{fig:cmdcl}
\end{figure}

Figure~\ref{fig:cmdcl} presents the CMD of cluster members from our refined sample and with a further constraint on a magnitude-dependent PM consistent with the cluster's bulk motion. The selection radii in the VPD as a function of magnitude bin were chosen as a compromise between excluding as many of the field stars with cluster-like PMs as possible, while retaining a statistical sample of bona-fide members. Apart from a few intermediate-age and old contaminants, the main sequence of the cluster appears to be well defined in the brighter part of the CMD, while field-star contamination becomes important for $\textrm{F090W} > -3.5$, due to larger PM errors.

Finally, we computed the intrinsic velocity dispersion of the cluster using the maximum-likelihood approach described in Sect.~4 of \citet{2022ApJ...934..150L}. We considered only bright, well-measured stars with $-6.5 < {\rm F090W} < -5$ within 25 arcsec from the center of the cluster, and with a PM within 0.091 mas yr$^{-1}$ (3.5 times the $\sigma$ of the magenta Gaussian in Fig.~\ref{fig:rotvpd}) from the bulk motion of the cluster. We find a velocity dispersion of $23.7 \pm 1.8$ $\mu$as yr$^{-1}$, i.e., $5.6 \pm 0.4$ km s$^{-1}$ assuming the cluster is at the same distance of the LMC \citep[49.5 kpc;][]{2019Natur.567..200P}. LMC clusters with masses similar to that of OGLE-CL LMC 407 \citep[$\sim$4000 $M_\odot$;][]{2012ApJ...751..122P} are expected to have velocity dispersions of the order of 1 km s$^{-1}$ \citep{2005ApJS..161..304M}. It is thus possible that the true dispersion of OGLE-CL LMC 407 is in fact lower, in which case the value of the velocity dispersion we computed can be interpreted as an upper limit on the possible presence of uncorrected systematic errors in either the first-epoch \hst or the second-epoch \jwst catalogs, or in both.

The intrinsic velocity dispersion of RGB stars implied by Fig.~\ref{fig:vpdoy} is $33.8 \pm 0.6$ km s$^{-1}$ ($0.144 \pm 0.003$ mas yr$^{-1}$). The distribution for the young populations is significantly more compact than this (but with significant substructure, per the preceding discussion), implying lower velocity dispersions. This is consistent with our general theoretical understanding of disk-heating mechanisms, which imply that old stars tend to have higher velocity dispersions than young stars. The values we report here are comparable to what is implied by line-of-sight velocities of LMC tracers \citep{2003MNRAS.339..701Z,2009IAUS..256...81V,2007ApJ...656L..61O,2014ApJ...781..121V}. Our findings are also consistent with those inferred from {\hst}-only PMs computed by \citet[see also their Fig.~15]{2021jwst.rept.7716A}. As a result of asymmetric drift, older populations have lower disk rotation than young populations, as has been verified observationally \citep{2014ApJ...781..121V,2021A&A...649A...7G}. The LMC is a ${\sim}10$ times less massive galaxy than the Milky Way \citep{2016ARA&A..54..529B,2019MNRAS.487.2685E}. So it also makes physical sense that the LMC velocity dispersion is several times lower than that of the Milky Way bulge, which has a central velocity dispersion $\sim$130--140 km s$^{-1}$ \citep{2019MNRAS.487.5188S,2018A&A...616A..83V}. 

It is thus intriguing that our results conflict with the large central velocity dispersion values reported for the LMC disk by \citet{2021A&A...649A...7G}. Our field is at 0.4 degrees from their adopted LMC center. At that distance, they report velocity dispersions for RGB stars and several young populations of order 70--80 km s$^{-1}$, well in excess of what is implied by Fig.~\ref{fig:vpdoy}. We do not know the cause of this discrepancy. Our samples have significantly smaller PM uncertainties than the \gaia DR3 PMs, so we may just be better resolving the kinematic structure near the crowded LMC center. It is possible that blending effects may have artificially inflated the scatter in the \gaia measurements near the dense LMC center. \citet{2021A&A...649A...7G} include a systemic error term of 0.01 mas yr$^{-1}$ in their analysis, but do not explicitly account for additional uncertainties introduced by blending and crowding. Rejection of stars based on the use of flags in the \gaia catalog will remove some of these, but maybe not all of them. Studies of runaway stars from the LMC 30~Dor region based on \gaia DR2 found that the PM scatter in the catalog in dense regions was significantly inflated \citep[i.e., implying unrealistic numbers of runaway stars;][]{2018A&A...619A..78L}. It will be of interest to reassess the velocity dispersions in the LMC disk based on future \gaia data releases that may include more sophisticated treatment of blended sources.

Overall, our astrometric studies clearly demonstrate the exquisite quality of our ePSF models and GD solutions, as well as of the overall astrometric and photometric potential of NIRISS, and that of \jwst in general. Moreover, the PMs for the science application presented here were obtained, for simplicity, by only using one twelfth of the available Commissioning data. A larger data set and more sophisticated techniques \citep[e.g., ][]{2014ApJ...797..115B,2021MNRAS.500.3213L,2022ApJ...934..150L} instead of simple global transformations between two frames could potentially improve the obtainable PM precision from Commissioning data. With further analyses it would not be out of the question to obtain PM uncertainties for the LMC approaching 1--2 km s$^{-1}$.

\section{Public release of our ePSF and GD models}\label{sec:release}

We make our library ePSF models and GD corrections publicly available to the community\footnote{The ePSF models and GD corrections in format \#1 are available at 
\href{https://www.stsci.edu/~jayander/JWST1PASS/LIB/PSFs/STDPSFs/NIRISS/}{STDPSFs} and \href{https://www.stsci.edu/~jayander/JWST1PASS/LIB/GDCs/STDGDCs/NIRISS/}{STDGDCs}, respectively. The GD corrections in format \#2 are available at \href{https://www.stsci.edu/~jayander/JWST1PASS/LIB/GDCs/OTHER/NIRISS/}{OTHERGDCs}.} as data-cube FITS files. The format of the ePSF models is described at the end of Sect.~\ref{sec:psf}. The GD-correction files are provided in two formats. The FITS files in format \#1 contain five layers. The value of each pixel in layer (1) gives the $x$ coordinate of each pixel in the 2048$\times$2048-pixel$^2$ chip into the GD-corrected reference frame. This value is provided only for the center of the pixel; the GD correction in any subpixel position should be derived using a bi-linear interpolation. Layer (2) is similar but for the $y$ coordinate. Layer (3) contains the pixel-area photometric corrections to be added to the magnitudes measured in the level-2 \texttt{\_cal} images. Layers (4) and (5) are the analogs of layers (1) and (2) but for the inverse GD correction, to map pixels from the GD-corrected frame into the raw NIRISS frame. We refer to Appendix~G of \citet{2022acs..rept....2A,2022wfc..rept....5A} for a detailed description of the GD-correction files in this format. The GD-correction FITS files in format \#2 are similar to those in format \#1, but they instead provide the actual values of the GD correction in layers (1) and (2), and again the pixel-area corrections in layer (3).

Recently, \citet{2022acs..rept....2A,2022wfc..rept....5A} released a FORTRAN routine, \texttt{hst1pass}, designed for ePSF-fitting of undersampled \hst images. \texttt{hst1pass} can provide coordinates in the raw system of a given \hst image, but it can also output GD-corrected positions. A version of \texttt{hst1pass} adapted to handle \jwst images, \texttt{jwst1pass}, has been used in the analysis described in this study and a preliminary version of this code is released along with this paper\footnote{The FORTRAN code is available at \href{https://www.stsci.edu/~jayander/JWST1PASS/CODE/JWST1PASS/}{CODE}.}
and will be described to the community in a future paper (Anderson et al., in preparation). Our ePSFs and GD corrections in format \#1 are designed to be used directly by \texttt{jwst1pass}, and users should be able to use them in their analysis of NIRISS images following the prescriptions in \citet{2022acs..rept....2A,2022wfc..rept....5A}.

Our ePSFs can potentially be used with other software packages, e.g., DOLPHOT \citep{2000PASP..112.1383D,2016ascl.soft08013D} or \texttt{photutils} \citep{larry_bradley_2022_6825092}, so long as these packages follow the same ePSF treatment and conventions described at the end of Sect.~\ref{sec:psf} or convert the ePSFs into a format that they can use. Similarly, our GD-correction FITS files in format \#2 can be easily used with any program that can handle the FITS standard, including \texttt{python}.

\section{Discussion and Conclusions}\label{sec:conc}

Proper motions have long been critical to our understanding of stellar systems and structures in the solar neighborhood. Starting in the early 2000's, \hst greatly advanced this field thanks to improved imagers, observations covering increasingly longer temporal baselines and new techniques. This made it possible to address for the first time the transverse kinematics of a wide range of stellar systems and structures throughout the entire Local Group \citep{2015IAUS..311....1V}, while also enabling new studies of stellar and exoplanetary systems in the Solar neighborhood \citep{2017MNRAS.470.1140B,2021MNRAS.501..911F}. \gaia further revolutionized this field starting in 2016, by combining high astrometric accuracy and well-calibrated systematics with all-sky coverage \citep{2021ARA&A..59...59B}. Despite the advances of \gaia, space observatories like \hst continue to be unique for a wide range of studies, even given their relatively small field of view.  Thanks to the larger mirror size and the ability to integrate for long times on specific fields, the same astrometric accuracy can be achieved at much fainter magnitudes. This offers many advantages, including the ability to accurately determine PMs further out in the Local Group \citep[e.g.,][]{2020ApJ...901...43S}. Moreover, the high spatial resolution makes it possible to resolve, e.g., the centers of star clusters that are too crowded for \gaia \citep{2010ApJ...710.1032A,2014ApJ...797..115B,2017ApJ...842....6B,2015ApJ...803...29W,2018ApJ...861...99L,2019ApJ...873..109L,2022ApJ...936..154W,2022ApJ...934..150L,2023ApJ...944...58L,2022ApJ...936..135S}. Finally, the combination of data from \hst and \gaia offers additional advantages 
\citep{2020A&A...633A..36M,2022ApJ...933...76D}, in that it allows the intrinsically differential nature of small-field astrometry to be calibrated to an absolute system.

This is the first in a series of papers by our \jwst Telescope Scientist Team to extend the astrometric legacy and lessons learned from \hst to {\jwst}. The {\jwst} observatory is uniquely suited for this, due to its high image quality and temporal stability \citep{2023arXiv230101779M}.  This offers several exciting new future opportunities, including: (a) the ability to study much fainter objects, owing to large mirror size of {\jwst} compared to {\hst}. In turn, this yields better statistics and accuracy in measurements of the mean and dispersion in kinematic properties, while also enabling PM studies of more distant objects; (b) the ability for PM studies of regions that are too extincted for \hst or \gaia, due to {\jwst}'s infrared sensitivity; and (c) the ability to extend the time baseline of imagery with space-based resolution to an era when {\hst} may no longer be available. All this will provide the opportunity to further advance the types of studies in which \hst and \gaia have already excelled, while also opening the door for entirely new studies and discoveries.

Here we have focused on developing the tools necessary to obtain high-precision astrometry and photometry with the NIRISS instrument of \jwst. We described in great detail the careful analyses we performed to construct accurate ePSF models and GD corrections, and the rigorous testing we put our products through. This has also allowed us to obtain new insights on the NIRISS detector performance and on \jwst in general. We make our ePSF models and GD corrections publicly available, together with a preliminary version of the dedicated software we are developing to analyze data taken with all \jwst's imagers, which will be described to the community in a future paper (Anderson et al., in preparation). Our combined efforts will provide to the community the tools necessary to achieve with {\jwst} the same outstanding results obtained for point sources that have been possible with \hst data.

As a scientific application and validation of our work, we studied the population-dependent PM kinematics of the stars in the standard \jwst calibration field near the center of the LMC disk. We determined the stellar PMs by combining second-epoch \jwst Commissioning data with a first-epoch \hst archival catalog based on data taken 16 years prior. For stars with apparent magnitudes around ${\it G} \sim 20$, the median PM uncertainty is $\sim$13 $\mu$as yr$^{-1}$, corresponding to 3.1 km s$^{-1}$ at the distance of the LMC. This is 1.2--1.8 times better than {\gaia} DR3 achieves for its best-measured stars (${\it G} \sim 9$--12), and 3--5 times better than what it achieves at $G \sim 16$, around the tip of the LMC RGB. 

These stunning accuracies are further validated by our ability to kinematically resolve tight substructures in the VPD of young and intermediate-age blue stars. We kinematically detect a known star cluster, OGLE-CL LMC 407. We measure for the first time its absolute PM and show how this differs from other blue LMC disk populations in the field. We also measure the cluster velocity dispersion to be $23.7 \pm 1.8$ $\mu$as yr$^{-1}$, corresponding to $5.6 \pm 0.4$ km s$^{-1}$. This provides an upper limit on the possible presence of any remaining uncorrected systematic errors. These results provide a testament to the overall quality of our \jwst NIRISS ePSF and GD-correction models. 

The RGB stars in the LMC field show no obvious kinematic substructures, and have a velocity dispersion of $33.8 \pm 0.6$ km s$^{-1}$ ($0.144 \pm 0.003$ mas yr$^{-1}$). This is comparable to results obtained from a large body of line-of-sight velocity studies, and reasonable by comparison to the central kinematics of such stars in the more massive Milky Way. However, this dispersion is 2--2.5  times less than what has been reported from studies of the \gaia DR3. This is the case for most of the younger blue populations in the field as well. We hypothesize that these discrepancies could be due to uncorrected increases in \gaia PM scatter due to crowding and blending near the dense LMC center.

In summary, our study successfully validates that \jwst is an extremely capable observatory for high-accuracy astrometry and PM studies in the Local Group. Unlike with \hst, where it took more than a decade for its full astrometric potential to be realized, this capability is available and accessible for the astronomical community with \jwst right from the start. We therefore anticipate that PM studies will become a key new component of \jwst's already vast arsenal of observational capabilities to further revolutionize astronomy over the coming decade. 

\section*{Acknowledgments}

The authors thank the anonymous referee for the thoughtful suggestions that improved the quality of our paper. This paper reports work carried out in the context of the \textit{JWST} Telescope Scientist Team\footnote{\href{https://www.stsci.edu/~marel/jwsttelsciteam.html}{https://www.stsci.edu/$\sim$marel/jwsttelsciteam.html}.} (PI: M.Mountain). Funding is provided to the team by NASA through grant 80NSSC20K0586.

NHA acknowledges support by the National Science Foundation Graduate Research Fellowship under Grant No. DGE1746891. DRL acknowledges research support by an appointment to the NASA Postdoctoral Program at the NASA Goddard Space Flight center, administered by Oak Ridge Associated Universities (ORAU) under contract with NASA. AG acknowledges support from the Robert R Shrock Graduate Fellowship. KKWH acknowledges funding from the Giacconi Fellowship at the Space Telescope Science Institute. JG acknowledges support from SERB through SERB Startup Research Grant SRG/2022/000727. RJM acknowledges support from NASA through the NASA Hubble Fellowship grant HST-HF2-51513.001 awarded by the Space Telescope Science Institute, which is operated by the Association of Universities for Research in Astronomy, Inc., for NASA, under contract NAS5-26555.

This research was pursued in collaboration with the HSTPROMO (High-resolution Space Telescope PROper MOtion) collaboration\footnote{\href{https://www.stsci.edu/~marel/hstpromo.html}{https://www.stsci.edu/$\sim$marel/hstpromo.html}.}, a set of projects aimed at improving our dynamical understanding of stars, clusters and galaxies in the nearby Universe through measurement and interpretation of proper motions from \textit{HST}, \textit{JWST}, \textit{Gaia}, and other space observatories. We thank the collaboration members for the sharing of their ideas and software.

Based on observations with the NASA/ESA/CSA \textit{JWST}, obtained at the Space Telescope Science Institute, which is operated by AURA, Inc., under NASA contract NAS 5-03127. Also based on observations with the NASA/ESA \textit{HST}, obtained at the Space Telescope Science Institute, which is operated by AURA, Inc., under NASA contract NAS 5-26555. This work has made use of data from the European Space Agency (ESA) mission {\it Gaia} (\url{https://www.cosmos.esa.int/gaia}), processed by the {\it Gaia} Data Processing and Analysis Consortium (DPAC, \url{https://www.cosmos.esa.int/web/gaia/dpac/consortium}). Funding for the DPAC has been provided by national institutions, in particular the institutions participating in the {\it Gaia} Multilateral Agreement. This research made use of \texttt{astropy}, a community-developed core \texttt{python} package for Astronomy \citep{astropy:2013, astropy:2018}.

\facilities{\jwst (NIRISS), \hst, \gaia}
\software{FORTRAN, \texttt{python}, \texttt{astropy}, \texttt{photutils}}

\appendix

\setcounter{table}{0}
\renewcommand{\thetable}{A\arabic{table}}

\setcounter{figure}{0}
\renewcommand{\thefigure}{A\arabic{figure}}

\begin{table*}[t!]
  \caption{Information about the processing of the \jwst data used in this work.}
  \centering
  \label{tab:context}
  \begin{tabular}{ccccc}
    \hline
    \hline
    Program ID & \texttt{CRDS\_CTX} & \texttt{CAL\_VER} & \texttt{SDP\_VER} & \texttt{PRD\_VER} \\
    1086 & jwst\_0937.pmap & 1.5.3 & 2022\_2a & PRDOPSSOC-055 \\
    1334 & jwst\_0874.pmap & 1.5.2 & 2022\_2  & PRDOPSSOC-052 \\
    1501 & jwst\_0977.pmap & 1.6.2 & 2022\_3  & PRDOPSSOC-057 \\
    \hline
  \end{tabular}
\end{table*}

\section{\jwst data processing information}\label{context}

Table~\ref{tab:context} provides information about the processing of the \jwst data used in this work. We report the CRSD context (\texttt{CRDS\_CTX}), the version of the calibration pipeline (\texttt{CAL\_VER}), the version number of the data processing software (\texttt{SDP\_VER}) and the Science and opertaion (S\&OC) Project Reference Database (S\&OC PRD) version number used in data processing (\texttt{PRD\_VER}).

\bibliography{NIRISS_Astrometry}{}
\bibliographystyle{aasjournal}

\allauthors

\vfill

{\footnotesize \noindent$^\star$ NSF Graduate Research Fellow~\\}
{\footnotesize \noindent$^\dagger$ Giacconi Fellow~\\}
{\footnotesize \noindent$^\ddagger$ NASA Postdoctoral Program Fellow~\\}
{\footnotesize \noindent$^\S$ NHFP Sagan Fellow}

\end{document}